\documentclass[12pt]{iopart}
\usepackage{cite}

\usepackage[utf8]{inputenc}
\usepackage[T1]{fontenc}
\usepackage[english]{babel}

\usepackage{graphicx}
\usepackage{subfigure}

\usepackage{amsmath}
\usepackage{amssymb}

\usepackage{bm}

\usepackage{units}
\usepackage[squaren]{SIunits}

\usepackage[version=3]{mhchem}

\usepackage{acronym}

\usepackage{hyperref}

\usepackage{mathtools}

\usepackage{tabularx,dcolumn,booktabs}

\usepackage{upgreek}

\usepackage{color}

\graphicspath{{./pics/}}

\newcommand{\sno}[1]{_\mathrm{#1}}
\newcommand{\no}[1]{\mathrm{#1}}

\begin{document}

\title[Rydberg systems in parallel electric and magnetic fields]{Rydberg systems in parallel electric and magnetic fields: an improved method for finding exceptional points}
%\date{\today}
\author{Matthias Feldmaier, J\"org Main, Frank Schweiner, Holger Cartarius and G\"unter Wunner}
\address{Institut f\"ur Theoretische Physik 1, Universit\"at
  Stuttgart, 70550 Stuttgart, Germany}

\begin{abstract}
Exceptional points are special parameter points in spectra of open quantum systems, at which resonance energies become degenerate and the associated eigenvectors coalesce. Typical examples are Rydberg systems in parallel electric and magnetic fields, for which we solve the Schrödinger equation in a complete basis to calculate the resonances and eigenvectors. Starting from an avoided crossing within the parameter-dependent spectra and using a two-dimensional matrix model, we develop an iterative algorithm to calculate the field strengths and resonance energies of exceptional points and to verify their basic properties. Additionally, we are able to visualise the wave functions of the degenerate states. We report the existence of various exceptional points. For the hydrogen atom these points are in an experimentally inaccessible regime of field strengths. However, excitons in cuprous oxide in parallel electric and magnetic fields, i.\,e., the corresponding hydrogen analogue in a solid state body, provide a suitable system, where the high-field regime can be reached at much smaller external fields and for which we propose an experiment to detect exceptional points.
\end{abstract}


\maketitle

% Abbreviations
\acrodef{EP}{exceptional point}
\acrodef{OM}{octagon method}

\section{Introduction}
\label{sec:introduction}
Rydberg systems in external fields are important examples of quantum systems which can be accessed theoretically via numerical calculations as well as experimentally in various cases (see Ref. \cite{gallagher1988rydberg} and references therein). A feature, found numerically, of the open quantum system of a hydrogen atom in \emph{crossed} electric and magnetic fields is the occurrence of \acp{EP} within the high-field regime~\cite{cartarius2007exceptional}. At these special points in the two-dimensional parameter space, spanned by the strengths of the external magnetic and electric field, not only two resonances become degenerate, but also the two corresponding eigenvectors coalesce~\cite{kato1976perturbation}. Encircling an \ac{EP} in parameter space leads to a typical exchange behaviour of the corresponding resonances within the complex energy plane~\cite{heiss2000repulsion}. \textcolor{black}{Here, isolated parameter values on a closed loop with small distances in-between are sufficient to observe the exchange behaviour of the eigenvalues (see, e.\,g., Ref.~\cite{dembowski2001experimental}). As shown in Ref.~\cite{cartarius2007exceptional} the complex eigenvalues can be extracted from the resonance spectra via the harmonic inversion method. Thus an adiabatic propagation of a wave packet (as, e.\,g., in Ref.~\cite{menke2016state}) is not required. The continuous connection of the eigenvalues in the complex energy plane leads to a clear and unambiguous proof of an \ac{EP}.} Theoretically, the occurrence of \acp{EP} has been shown for, e.\,g., atomic spectra \cite{magunov1999strong,magunov2001laser,latinne1995laser,cartarius2007exceptional} and molecular spectra \cite{lefebvre2009resonance}, optical waveguides \cite{klaiman2008visualization} or resonators \cite{wiersig2008asymmetric,wiersig2014enhancing}, and they have been found experimentally, e.\,g., in microwave cavities \cite{dembowski2001experimental,dietz2011exceptional,bittner2014scattering}, electronic circuits \cite{stehmann2004observation}, metamaterials \cite{lawrence2014manifestation}, and exciton-polariton resonances \cite{gao2015observation}. \textcolor{black}{Exceptional points in open quantum systems can be used to transfer population between the related resonances as has been done, e.\,g., for vibrational modes of the $\no{H}_2^+$ ion and the $\no{Na}_2$ molecule \cite{jaouadi2013signatures,atabek2011proposal}. In addition, a laser-controlled rotational cooling of $\no{Na}_2$ could be realised based on exceptional points \cite{kokoouline2013laser}. In a completely different application, \acp{EP} have shown to be extremely important for the enhancement of the sensitivity of optical detectors \cite{wiersig2014enhancing,li2016pt}. \acp{EP} may also be used to generate Majorana bound states in superconductors \cite{san2014majorana}.}

An experimental validation of the predicted \acp{EP} for the hydrogen atom seems currently out of reach because the strength of the external magnetic field needed to access the high-field regime is in the order of several ten to hundreds of Tesla~\cite{cartarius2007exceptional}, which yet cannot be realised experimentally. On the other hand, due to computational limits, precise numerical calculations for the hydrogen atom are only possible at high fields, which are strong enough to lead to the coalescence of levels at low energies. The ability to search for \acp{EP} numerically at higher levels and therefore at lower fields within the spectrum of the hydrogen atom is limited by the increasing number of states to be taken into account. This short-coming may be overcome with the help of recent high resolution absorption experiments with Rydberg excitons in cuprous oxide ($\no{Cu}_2\no{O}$), which exhibit a hydrogen-like spectrum up to a principal quantum number of $n=25$~\cite{kazimierczuk2014giant, assmann}. Here, the common description of excitons similar to a hydrogen atom with a Coulomb interaction between a negatively charged electron and a positively charged hole (see, e.\,g., Ref.~\cite{roessler2009solid}) seems to be appropriate. In $\no{Cu}_2\no{O}$ the strengths of the external fields to enter the high-field regime are much smaller compared to the hydrogen atom~\cite{agekyan1977spectroscopic}. Hence, it provides a system that is appropriate to check theoretical predictions experimentally.

In this paper we will demonstrate the occurrence of \acp{EP} for Rydberg systems in \emph{parallel} electric and magnetic fields. We verify several \acp{EP} by the typical exchange behaviour of the associated eigenvectors \cite{heiss2000repulsion}, and then try to locate the precise position of these \acp{EP} within the parameter space spanned by the strengths of the external magnetic and electric field. Using the recently proposed 3-point method of Uzdin and Lefebvre \cite{uzdin2010finding} we are in principle able to locate these positions, however, with a huge amount of computational effort. Therefore, we develop a new and much less expensive method, the \ac{OM}, which is based on a two-dimensional matrix model to describe the vicinity of the two states forming an \ac{EP}. Using the \ac{OM}, we are able to calculate the precise position of a variety of \acp{EP} within the cylindrically symmetric system of a Rydberg atom in parallel electric and magnetic fields. Additionally, the \ac{OM} allows us to simulate the exchange behaviour of the two associated resonances and their paths in the complex energy plane while encircling an \ac{EP} in parameter space without further time-consuming quantum-mechanical calculations. For Rydberg excitons in $\no{Cu}_2\no{O}$ these \acp{EP} are located in an experimentally accessible regime of the external field strengths. Therefore we propose an experiment to verify our theoretical predictions by measurements of photoabsorption spectra in cuprous oxide.

A second purpose of this paper is to visualise the two states associated with an \ac{EP}. At the \ac{EP} the two probability distributions of the (numerically) degenerate states do not coincide exactly as a result of limited computational accuracy. They differ by various sets of structured lines. These lines show the same exchange behaviour when encircling the \ac{EP} in parameter space as has already been demonstrated experimentally, e.\,g., for microwave cavities in Ref.~\cite{dembowski2001experimental}. By averaging the eigenvectors of the associated states we compute the probability distribution of the degenerate states directly at the \ac{EP}.

The paper is organised as follows: In Sec.~\ref{sec:methods} we present the modelling of Rydberg systems in parallel electric and magnetic fields and discuss the similarities and differences between hydrogen and cuprous oxide. In subsection~\ref{sec:computation_of_eigenstates} follows the theory for the computation of eigenstates. In subsection~\ref{sec:octagon_method} we use a two-dimensional matrix model to develop the \ac{OM} as an iterative algorithm to find the exact position of an \ac{EP} in the parameter space as well as its precise complex resonance energy. The next subsection \ref{sec:verification} deals with the question how the points of convergence of this new method can be verified as \acp{EP} without unnecessary time-consuming calculations. In section~\ref{sec:results} all results are presented. The \acp{EP} found are listed and discussed, and we also visualise the associated states directly at the \ac{EP}.

\section{Methods and theoretical background}
\label{sec:methods}
% According to the model for the hydrogen atom in crossed electric and magnetic fields in Ref.~\cite{cartarius2007exceptional},
The Hamiltonian of the hydrogen atom in parallel electric ($F$) and magnetic ($B$) fields, which are both orientated along the $z$-axis, reads
\begin{equation}
H\sno{hyd} = \frac{\mathbf{p}^2}{2m_0} - \frac{e^2}{4\pi\varepsilon_0}\,\frac{1}{r} + \frac{e\,B}{2m_0}\,L_z + \frac{e^2\,B^2}{8m_0}\left(x^2+y^2\right) + e\,F\,z,
\label{ham_hyd}
\end{equation}
where $e$ denotes the elementary charge, $\mathbf{p}$ the relative momentum, and $r = \sqrt{x^2 + y^2 + z^2}$ is the distance between the electron with mass $m_0$ in free space and the proton with approximately infinite mass. The vacuum permittivity is denoted by $\varepsilon_0$, and $L_z$ is the $z$-component of the angular momentum operator. 

To set up a Hamiltonian similar to Eq.~\eqref{ham_hyd} for a Rydberg exciton in $\no{Cu}_2\no{O}$ we use the simple band model \cite{roessler2009solid}. The external fields are included via minimal substitution. We introduce relative and centre-of-mass coordinates and set the pseudomomentum of the centre of mass motion to zero. \textcolor{black}{The resulting Hamiltonian for a Rydberg exciton in $\no{Cu}_2\no{O}$ reads \cite{schmelcher1992regularity}
\begin{equation}
H\sno{ex} = \frac{\mathbf{p}^2}{2\mu} - \frac{e^2}{4\pi\varepsilon_0\,\varepsilon\sno{r}}\,\frac{1}{r} + \frac{e\,B}{2\mu}\,\frac{m\sno{h} - m\sno{e}}{m\sno{h}+m\sno{e}}\,L_z + \frac{e^2\,B^2}{8\mu}\left(x^2 + y^2\right) + e\,F\,z
\label{ham_ex}
\end{equation}
}
with the effective masses $m\sno{e} = 0.99m_0$ for the electron, $m\sno{h} = 0.62m_0$ for the hole, and $\mu = 0.38m_0$ being the reduced mass \cite{sasaki1973magneto}. Note that in Eq.~\eqref{ham_ex} we have not included a band gap energy. Therefore, the energies need to be corrected by an  offset of $E\sno{gap} = 2.17208\,\no{eV}$~\cite{kazimierczuk2014giant}. 

For reasons of symmetry due to the parallel fields, the angular momentum is a good quantum number\footnote{For $\no{Cu}_2\no{O}$ this is also true in the simple band model, since the cubic group is a group of very high symmetry \cite{roessler2009solid,thewes2015observation}.}; so the angular momentum operator $L_z$ can be replaced by its quantum number $m$. Hence, the corresponding paramagnetic term $H\sno{P}$ (the third term in Eq.~\eqref{ham_ex}) describes a $B$-dependent shift of the zero point energy. For this reason we can neglect this term in the following and treat only the Hamiltonian $H' = H- H\sno{P}$. To reduce the two equations \eqref{ham_hyd} and \eqref{ham_ex} to the same form we introduce appropriate units with material dependent constants $B_0$ and $F_0$, where $\gamma = B/B_0$ is the reduced magnetic flux density and $f=F/F_0$ the reduced strength of the electric field (see \ref{app_hartree}). Consequently, the Hamiltonian of relative motion for a hydrogen-like system in parallel electric and magnetic fields reads
\begin{equation}
H' = \frac{1}{2}\mathbf{p}\,^2 - \frac{1}{r} + \frac{1}{8}\gamma^2\,\left(x^2+y^2\right) + f\,z,
\label{hamiltonian_hartree}
\end{equation}
Hence, Eq.~\eqref{hamiltonian_hartree} becomes independent of the material parameters of the system (atomic or solid state). The respective properties are absorbed in the constants $B_0$ and $F_0$. For the hydrogen atom holds $B_0 = 2.350\,517\times 10^{5}\,\no{T}$ and $F_0 = 5.142\,206\times 10^{11}\,\frac{\no{V}}{\no{m}}$ and for $\no{Cu}_2\no{O}$ we obtain $B_0 = 603.4\,\no{T}$ and $F_0 = 1.760\times 10^{8}\,\frac{\no{V}}{\no{m}}$ (see \ref{app_hartree}).

\subsection{Computation of eigenstates}
\label{sec:computation_of_eigenstates}
\textcolor{black}{Due to the electric fields and the complex scaling we obtain resonances as eigenstates of the non-Hermitian Hamiltonian~\eqref{hamiltonian_hartree}.} In contrast to bound states, which have a real energy and infinite lifetime, the energies of the decaying resonance states become complex, whereupon the imaginary part describes the width of the resonance or its inverse lifetime~\cite{moiseyev1998quantum}. Resonances can be introduced by non-Hermitian operators using the complex rotation method~\cite{cartarius2007exceptional, moiseyev1998quantum, delande1991positive}.\\
To solve Eq.~\eqref{hamiltonian_hartree} it is useful to transform it to dilated semiparabolic coordinates~\cite{cartarius2009signatures, main1994rydberg}, 
\begin{equation}
\mu = \frac{\mu\sno{r}}{b} = \frac{1}{b}\,\sqrt{r+z},\quad \nu = \frac{\nu\sno{r}}{b} = \frac{1}{b}\,\sqrt{r-z},\quad \varphi = \arctan \left(\frac{y}{x}\right),
\label{semiparabolic}
\end{equation}
where $b \rightarrow |b|\,\no{exp}(\no{i}\alpha)$ is the complex dilation parameter that induces a complex rotation of the real semiparabolic coordinates $(\mu\sno{r}, \nu\sno{r}, \varphi)$. In dilated semiparabolic coordinates the Schrödinger equation of the Hamiltonian~\eqref{hamiltonian_hartree} reads
\begin{equation}
\left(2H_0 - 4b^2 + \frac{1}{4}b^8\,\gamma^2\,(\mu^4\,\nu^2 + \mu^2\,\nu^4) + b^6\,f\,(\mu^4-\nu^4)\right) \, \Psi = \lambda\,(\mu^2+\nu^2)\,\Psi,
\label{generalized}
\end{equation}
with the generalised eigenvalue $\lambda = 1 + 2b^4\,E'$. Here $E'$ is the energy of the respective quantum state and corresponds to the Hamiltonian $H'$ in Eq.~\eqref{hamiltonian_hartree}. The Hamiltonian $H_0 = H_\mu + H_\nu$ in Eq.~\eqref{generalized} is the sum of two two-dimensional harmonic oscillators $H_\rho$ with $\rho \in \{\mu,\nu\}$ to be taken as radial coordinate,
\begin{equation}
H_\rho = \left(-\frac{1}{2}\Delta_\rho + \frac{1}{2}\rho^2 \right), \quad \Delta_\rho = \frac{1}{\rho}\,\frac{\partial}{\partial \rho}\,\rho\,\frac{\partial}{\partial \rho} - \frac{m^2}{\rho^2},
\end{equation}
with $m$ being the magnetic quantum number. Hence, for the matrix representation of Eq.~\eqref{generalized} the eigenstates $\left|n_\rho, m\right>$ of the two-dimensional harmonic oscillator represent an appropriate basis~\cite{messiah_qm1}. Since the angle $\varphi$ in $H_\mu$ and $H_\nu$ is the same, just one fixed angular quantum number $m$ is needed to construct the basis 
\begin{equation}
\left|n_\mu, n_\nu, m\right> = \left|n_\mu , m\right> \otimes \left|n_\nu, m\right>
\label{total_basis}
\end{equation}
of the total system (we use $m=0$ in the following calculations, which yields $H = H'$ and $E = E'$). The relations needed to calculate the matrix representation of Eq.~\eqref{generalized} within the  basis \eqref{total_basis} can be found in Refs.~\cite{englefield1972group, feldmaier2015untersuchung}. Its diagonalization is done using the \textit{IRAM} method of the \textit{ARPACK} package \cite{lehoucq1998arpack}, where we consider only states up to a maximum quantum number of $n\sno{max} = n_\mu + n_\nu$. The value $n\sno{max} = 90$ has turned out to be sufficient in the following calculations. Convergence is assured by choosing a proper value for the parameter $b$ of the dilated semiparabolic coordinates~\cite{muller1994scars}. The position space representation of the basis~\eqref{total_basis} is given by (see, e.\,g., Ref.~\cite{feldmaier2015untersuchung} for a derivation)
\begin{align}
\begin{aligned}
\Psi_{n_\mu, n_\nu, m}(\mu, \nu, \varphi) &= \sqrt{\frac{n_\mu!\,n_\nu!}{(n_\mu + |m|)!\,(n_\nu + |m|)!}}\,\sqrt{\frac{2}{\pi}}\,f_{n_\mu, m}(\mu)\,f_{n_\nu, m}(\nu)\,\no{e}^{im\,\varphi},\\
f_{n_\rho, m}(\rho) &= \no{e}^{-\frac{\rho^2}{2}}\,\rho^{|m|}\,L^{|m|}_{n_\rho}(\rho^2)\quad\mathrm{for}\quad \rho \in \{\mu, \nu\},
\end{aligned}
\label{wavefunction}
\end{align}
with the generalised Laguerre polynomials $L_n^{m}(\rho)$. Note that the correct inner product for the wave functions of the non-Hermitian Hamiltonian~\eqref{generalized} obtained with the complex scaling approach is achieved by complex conjugating only the intrinsically complex parts ($\no{exp}(\no{i}m\,\varphi) \rightarrow \no{exp}(-\no{i}m\,\varphi)$) and not that originating from the complex parameter $b$ \cite{moiseyev1998quantum}. We use this complex conjugation throughout this article.
% Note that to calculate the complex conjugate $\Psi^*_{N_\mu, N_\nu, m}(\mu, \nu, \varphi)$ only the intrinsic complex part $\exp(\no{i}m\,\varphi)$ of Eq.~\eqref{wavefunction} has to be replaced by $\exp(-\no{i}m\,\varphi)$. Although the coordinates $\mu$ and $\nu$ become complex by complex rotation, they must not be conjugated~\cite{moiseyev1998quantum}. 

In dilated semiparabolic coordinates the spatial wave functions of a state with energy $E_i$ read
\begin{subequations}
\begin{align}
\Psi_i(\mu, \nu, \varphi) &= \sum_{n_\mu, n_\nu} c_{i, n_\mu, n_\nu, m}\, \Psi_{n_\mu, n_\nu, m}(\mu, \nu, \varphi),\\
\Psi_i^*(\mu, \nu, \varphi) &= \sum_{n_\mu, n_\nu} c_{i, n_\mu, n_\nu, m}\, \Psi^*_{n_\mu, n_\nu, m}(\mu, \nu, \varphi)
\end{align}
\label{wave_functions_semiparabolic}%
\end{subequations}
with the expansion coefficients $c_{i, n_\mu, n_\nu, m}$ of the associated eigenvector obtained by the matrix diagonalization of Eq.~\eqref{generalized}.

\subsection{Octagon method for the exact localisation of \acp{EP}}
\label{sec:octagon_method}
The purpose of this article is to verify the existence of \acp{EP} for Rydberg systems in parallel external fields and to locate their precise position in the parameter space spanned by the strength $f$ of the electric and $\gamma$ of the magnetic field. A recently proposed iterative algorithm to calculate the exact position of an \ac{EP} in parameter space is the 3-point method of Uzdin and Lefebvre~\cite{uzdin2010finding}. It converges if the initial parameters are chosen close enough to the actual position of the \ac{EP}.
To find initial parameters for an \ac{EP} in parallel electric and magnetic fields we make use of the fact that level repulsion is associated with the occurrence of \acp{EP}~\cite{heiss2000repulsion}. Therefore, the parameters of an avoided crossing within the energy spectrum provide a good starting point to find an \ac{EP}. 

By applying the 3-point method to the system of an exciton in parallel electric and magnetic fields it turns out that the initial parameters of an avoided crossing are generally not close enough to the actual position of an \ac{EP} to achieve a convergent behaviour; so further highly expensive numerical calculations are needed~\cite{feldmaier2015untersuchung}. In this regard, we develop an improved method, which is able to converge from the initial parameters of an avoided crossing to the precise position of an \ac{EP}. To this aim, the basic properties of an \ac{EP} have to be taken into account.

\textcolor{black}{At an \ac{EP} not only the two resonances become degenerate, but also the two corresponding eigenvectors coalesce~\cite{kato1976perturbation}}. If an \ac{EP} is fully encircled in parameter space, its two related resonances exchange their position in the complex energy plane~\cite{heiss2000repulsion}. In a local vicinity around the \ac{EP} the two related states can be described using a two-dimensional matrix $\bm{M}$ (see, e.\,g., Ref.~\cite{heiss2000repulsion}). Any coupling to other states is not taken into account here. We assume the elements $M_{kl}$ of $\bm{M}$ with $k,l \in \left\{1,2\right\}$ to be linear expansions in the strengths $\gamma$ and $f$ of the external fields to describe their influence on the states in the vicinity of a centre-point ($\gamma_0, f_0$) \cite{cartarius2009exceptional}
\begin{equation}
M_{kl} = a_{kl}^{(0)} + a_{kl}^{(\gamma)}\,(\gamma - \gamma_0) + a_{kl}^{(f)}\,(f-f_0).
% \bm{M} = \begin{pmatrix} a_0 + a_\gamma\,(\gamma - \gamma_0) + a_f\,(f-f_0) & b_0 + b_\gamma\,(\gamma-\gamma_0) + b_f\,(f-f_0)\\ b_0 + b_\gamma\,(\gamma-\gamma_0) + b_f\,(f-f_0) & c_0 + c_\gamma\,(\gamma-\gamma_0)+ c_f\,(f-f_0) \end{pmatrix}.
\label{2d_matrix}
\end{equation}
% Note that $\bm{M}$ keeps the complex-symmetric structure of the system described by Eq.~\eqref{generalized}.
The eigenvalues $E_i$ with $i \in \{1, 2\}$ of $\bm{M}$ then fulfil
\begin{subequations}
\begin{gather}
\begin{multlined}
\kappa \equiv E_1 + E_2 = \no{tr}(\bm{M}) =  A + B\,(\gamma - \gamma_0) + C\,(f-f_0),\quad \quad\qquad \qquad~~~~~
\end{multlined} 
\\
\begin{multlined}
\eta \equiv (E_1 - E_2)^2 = \no{tr}^2(\bm{M}) - 4\, \no{det}( \bm{M}) = D + E\,(\gamma-\gamma_0) + F\,(f-f_0)\\ 
+ G\,(\gamma-\gamma_0)^2 + H\,(\gamma - \gamma_0)\,(f-f_0) + I\,(f-f_0)^2,\quad~~
\label{2dmatrixentwicklung}
\end{multlined}
\end{gather}\label{2dmatrixalles}\noindent
\end{subequations}
with complex coefficients $A$ to $I$. In contrast to the 3-point method of Uzdin and Lefebvre \cite{uzdin2010finding} our approach includes additional quadratic terms in Eq.~\eqref{2dmatrixentwicklung}.
% Neglecting the quadratic terms of Eq.~\eqref{2dmatrixentwicklung}, one would arrive at the 3-point method of Uzdin and Lefebvre~\cite{uzdin2010finding}. 
% Supplying a starting point close enough to the actual position ($\gamma\sno{EP}, f\sno{EP}$) of the \ac{EP}, the 3-point method provides an iterative algorithm converging to the \ac{EP}'s position.
For the Rydberg system in parallel electric and magnetic fields the 3-point method turned out to be very sensitive to the initial conditions. To obtain convergence, one has to start very close to ($\gamma\sno{EP}, f\sno{EP}$) and huge numerical effort is necessary to get that close~\cite{feldmaier2015untersuchung}.
\begin{figure}[t]
\centering
\includegraphics[width=0.5\columnwidth]{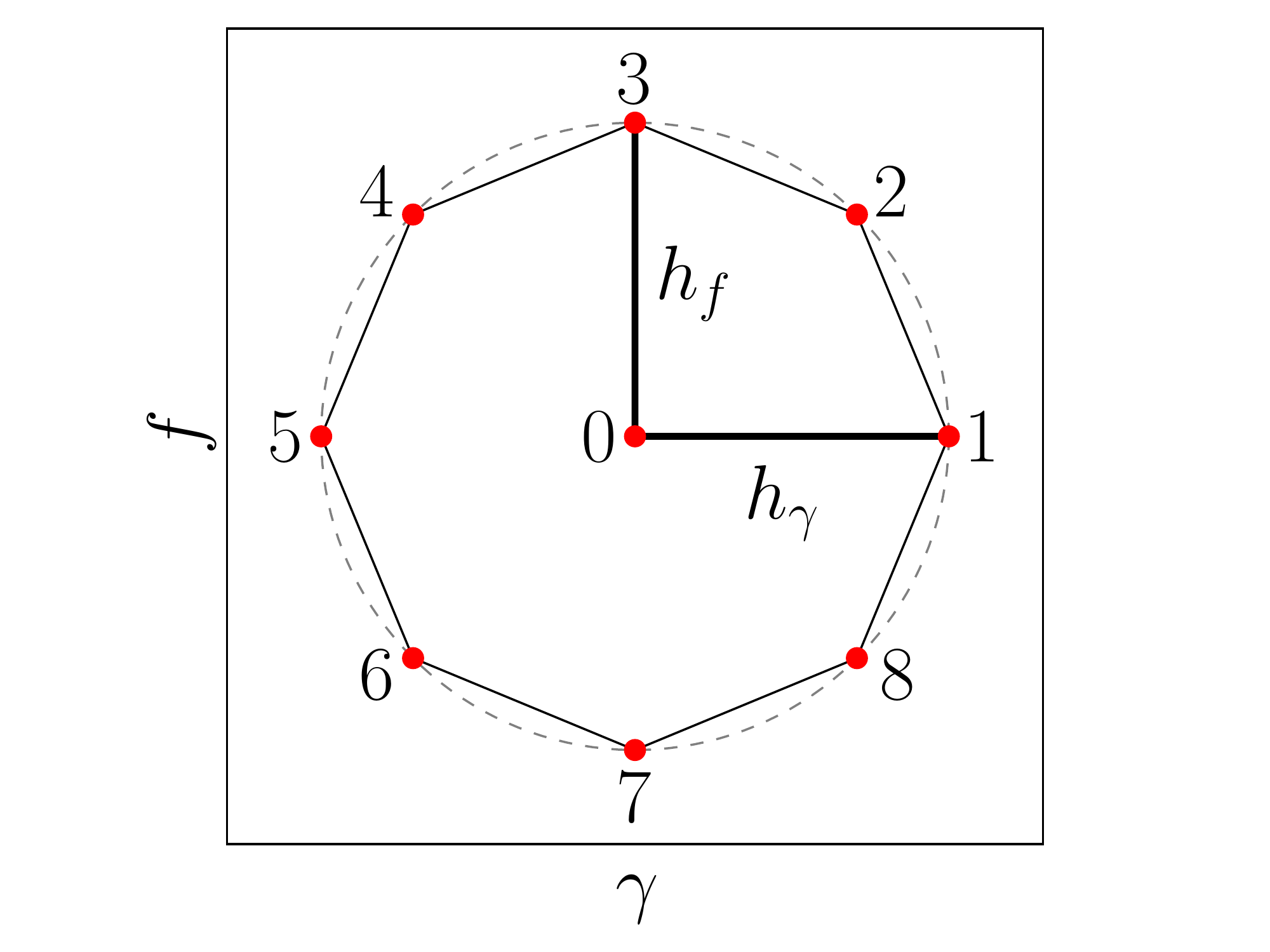}
\caption{Illustration of the octagon method. Nine points are arranged in the $(\gamma, f)$ parameter space, eight at the corners of an octagon and one at its centre. At each point $i \in \{0, 8\}$ the generalised eigenvalue equation \eqref{generalized} has to be solved to obtain $\kappa_i = E_{1,i} + E_{2,i}$ and $\eta_i \equiv (E_{1,i} - E_{2,i})^2$ for the two eigenvalues $E_1$ and $E_2$ that become degenerate at the \ac{EP}.}\label{fig:oktagon}%
\end{figure}
According to Ref.~\cite{uzdin2010finding}, \acp{EP} appear in the vicinity of an avoided crossing of two states; so it would be useful to have a method that converges when using the rough initial parameters of such an avoided crossing. The inclusion of the quadratic terms in Eq.~\eqref{2dmatrixentwicklung} leads to the \ac{OM}, which can be visualised by means of Fig.~\ref{fig:oktagon}. To get simple relations for the coefficients the \ac{OM} uses nine points in the ($\gamma, f$) parameter space; eight at the corners of an octagon and one at the centre to compute the coefficients $A$ to $I$ in Eqs.~\eqref{2dmatrixalles}. Solving Eq.~\eqref{generalized} at every point $i$ in Fig.~\ref{fig:oktagon} the two eigenvalues $E_{1,i}$ and $E_{2,i}$ associated with the \ac{EP} need to be calculated. The sum $\kappa_i = E_{1,i} + E_{2,i}$ and the squared difference $\eta_i \equiv (E_{1,i} - E_{2,i})^2$ can be used to determine all coefficients of Eq.~\eqref{2dmatrixalles}, i.\,e.,
% $D$ corresponds to the value $\eta_0$ at ($\gamma_0, f_0$). The other coefficients can be calculated by first and second order derivatives
\begin{align}
\begin{aligned}
A &= \kappa_0, &B &= \frac{\kappa_1 - \kappa_5}{2h_\gamma},\\
C &= \frac{\kappa_3 - \kappa_7}{2h_f}, &D &= \eta_0,\\
E &= \frac{\eta_1 - \eta_5}{2h_\gamma}, &F &= \frac{\eta_3 - \eta_7}{2h_f}, \\ 
G &= \frac{\eta_1 + \eta_5 - 2\eta_0}{2h_\gamma^2}, &I &= \frac{\eta_3 + \eta_7 - 2\eta_0}{2h_f^2},\\
H &= \frac{\eta_2 - \eta_4 + \eta_6 - \eta_8}{2h_\gamma\,h_f}.
\end{aligned}
\label{all_coeffs}
\end{align}
% \begin{itemize}
% \item The coefficient $D$ corresponds to the value $\eta_0$ at ($\gamma_0, f_0$).
% \item $E$ and $F$ are given by first numerical derivations
%   \begin{equation}
%     E = \frac{\eta_1 - \eta_5}{2h_\gamma}, \quad F= \frac{\eta_3 - \eta_7}{2h_f}.
%   \end{equation}
% \item $G$ and $I$ can be calculated using second numerical derivations
%   \begin{equation}
%     G = \frac{\eta_1 + \eta_5 - 2\eta_0}{2h_\gamma^2}, \quad I = \frac{\eta_3 + \eta_7 - 2\eta_0}{2h_f^2}.
%   \end{equation}
% \item To calculate the mixed term's coefficient $H$, four points are necessary
%   \begin{equation}
%     H = \frac{\eta_2 - \eta_4 + \eta_6 - \eta_8}{2h_\gamma\,h_f}.
%   \end{equation}
% \end{itemize}
Having calculated the coefficients~\eqref{all_coeffs} we can make an estimation for the position $(\gamma\sno{EP}, f\sno{EP})$ of the \ac{EP} by setting the left hand side of Eq.~\eqref{2dmatrixentwicklung} to zero, which is the condition for the degeneracy of the two eigenvalues. With the abbreviations $x \equiv (\gamma\sno{EP} - \gamma_0)$ and $y \equiv (f\sno{EP} - f_0)$ we obtain
\begin{equation}
0 = D + E\,x + F\,y + G\,x^2 + H\,x\,y + I\,y^2.
\label{null}
\end{equation}
To find the \ac{EP}, both the real and the imaginary part of the polynomial in Eq.~\eqref{null} must vanish. The resulting system of equations has in general four complex roots $(x,y)$. Only one of these four roots is indeed an estimation for $(\gamma\sno{EP}, f\sno{EP})$. The other three (possibly complex) roots arise due to the mathematical structure of approximating the squared energy difference in Eq.~\eqref{2dmatrixentwicklung} up to the second order in $\gamma$ and $f$, and therefore have no physical relevance. The formulas needed to calculate the four roots of Eq.~\eqref{null} as well as a method to choose the physically correct one are presented in \ref{app_solution}.

The position estimate $(\gamma\sno{EP}, f\sno{EP})$ can be taken as centre-point $(\gamma_0,f_0)$ of a new octagon in parameter space to calculate the eigenvalues of Eq.~\eqref{generalized}. Afterwards, a new set of coefficients~\eqref{all_coeffs} can be determined and a new estimation for $(\gamma\sno{EP}, f\sno{EP})$ can be made. Continuing this procedure step-by-step allows us to develop an iterative algorithm, which converges to the true position of the \ac{EP},
\begin{align}
\begin{aligned}
\gamma_0^{(n+1)} &= \gamma\sno{EP}^{(n)}; \quad \gamma\sno{EP} = \lim_{n\rightarrow \infty} \gamma_0^{(n)},\\
f_0^{(n+1)} &= f\sno{EP}^{(n)}; \quad f\sno{EP} = \lim_{n\rightarrow \infty} f_0^{(n)}.
\end{aligned}
\end{align}
Here, $(\gamma_0^{(n)}, f_0^{(n)})$ denotes the centre of the octagon during the $n$-th iteration step and $\gamma\sno{EP}^{(n)} = \gamma_0^{(n)} + x$ and $f\sno{EP}^{(n)} = f_0^{(n)} + y$ are the corresponding estimates for the position of the \ac{EP} obtained by solving for $x$ and $y$ in Eq.~\eqref{null}. 
 
% A more detailed description and additional steps of calculation can be found in Ref.~\cite{feldmaier2015untersuchung}.

\subsection{Verification of \acp{EP} found}
\label{sec:verification}
Once the \ac{OM} has converged to a specific point in the $(\gamma, f)$-parameter space, one has to prove that this point actually is an \ac{EP}. There are two possibilities for this verification. For the first one an \ac{EP} is fully encircled in parameter space, so the corresponding resonances will undergo an exchange of their position in the complex energy plane~\cite{kato1976perturbation}. To be able to sort the calculated resonances properly, Eq.~\eqref{generalized} has to be solved successively in small steps on the circle around the \ac{EP}. Hence, the computational effort is very high. By using the \ac{OM} this effort can be reduced considerably. Having obtained the coefficients $A$ to $I$ by solving Eq.~\eqref{generalized} numerically only at the nine points of the octagon in Fig.~\ref{fig:oktagon} the values of $\kappa$ and $\eta$ of Eq.~\eqref{2dmatrixalles} are well known for all sets of parameters $(\gamma, f)$, which lie in the vicinity of $(\gamma_0, f_0)$ where the approximations \eqref{2dmatrixalles} hold. Thus, for any point ($\gamma, f$) on the circle circumscribing the octagon (provided that the radius is small enough), the two resonances $E_1$ and $E_2$ can be calculated using Eq.~\eqref{2dmatrixalles} without further time-consuming diagonalization of Eq.~\eqref{generalized},
\begin{equation}
E_{1,2} = \frac{\kappa(\gamma, f) \pm \sqrt{\eta(\gamma, f)}}{2}.%,\quad E_2 = \frac{\kappa - \sqrt{\eta}}{2}.
\label{resonances_fit}
\end{equation}
The arising paths in the complex energy plane can now be visually checked for the exchange behaviour of the corresponding resonances.

Another possibility is the calculation of a winding number that allows for a clear statement about the existence of an \ac{EP} without the need to visually check the paths of exchanging resonances in the complex energy plane. Here again, the circle around the octagon, $\gamma(\varphi) = \gamma_0 + h_\gamma\,\cos(\varphi)$, $f(\varphi) = f_0 + h_f\,\sin(\varphi)$, is discretized in $n$ steps, which are denoted by respective angles $\varphi_i$ with $i \in \{0,\ldots, n\}$ as in the previous method. At each point $(\gamma(\varphi_i), f(\varphi_i))$ of the discretized circle the squared energy difference $\eta(\varphi_i)$ can be calculated easily using  Eq.~\eqref{2dmatrixentwicklung} with the known coefficients $D$ to $I$. The result is a closed curve for $\eta$ within the complex $\eta$-plane. If this $\eta$-curve encircles the origin of coordinates, an \ac{EP} is located within the circle in the $(\gamma, f)$-space~\cite{feldmaier2015untersuchung}. This can be tested numerically using the residue theorem by which we calculate the winding number of the $\eta$-curve for a function with a pole at the origin of coordinates with a discretized formula
\begin{equation}
W_{\eta=0} = \frac{1}{2\pi\no{i}}\,\sum_{i=1}^n\frac{\eta(\varphi_{i+1}) - \eta(\varphi_{i-1})}{2\eta(\varphi_i)}.
\label{winding_number}
\end{equation}
In case of $W_{\eta=0} = 1$ at a sufficiently high $n$ the resonances \eqref{resonances_fit} show the typical exchange behaviour of an \ac{EP}, which is located within the octagon-circle in the $(\gamma, f)$-space.

\section{Results and discussion}
\label{sec:results}
\begin{figure}[t]
\centering
\includegraphics[width=0.49\columnwidth]{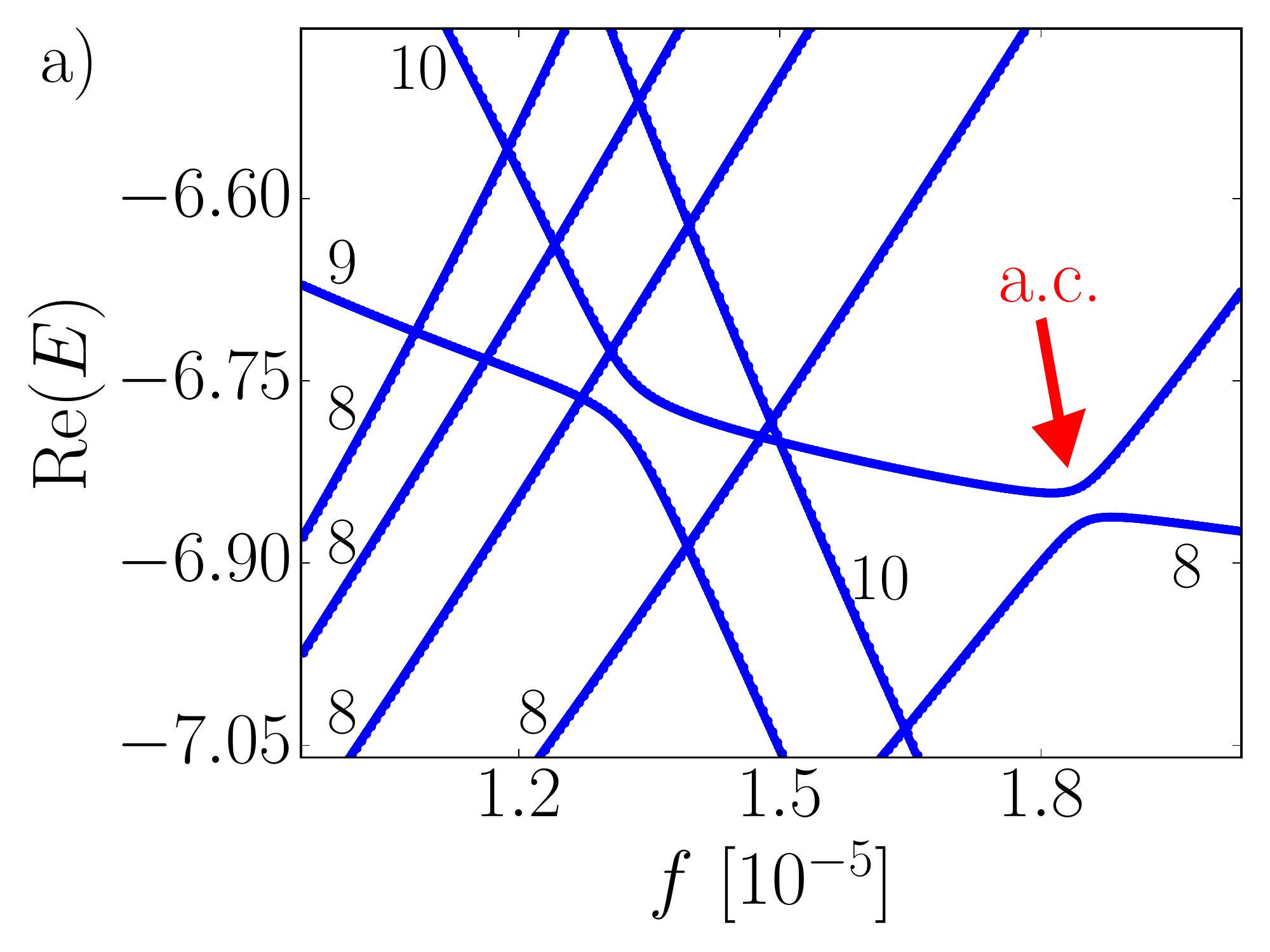}
\includegraphics[width=0.49\columnwidth]{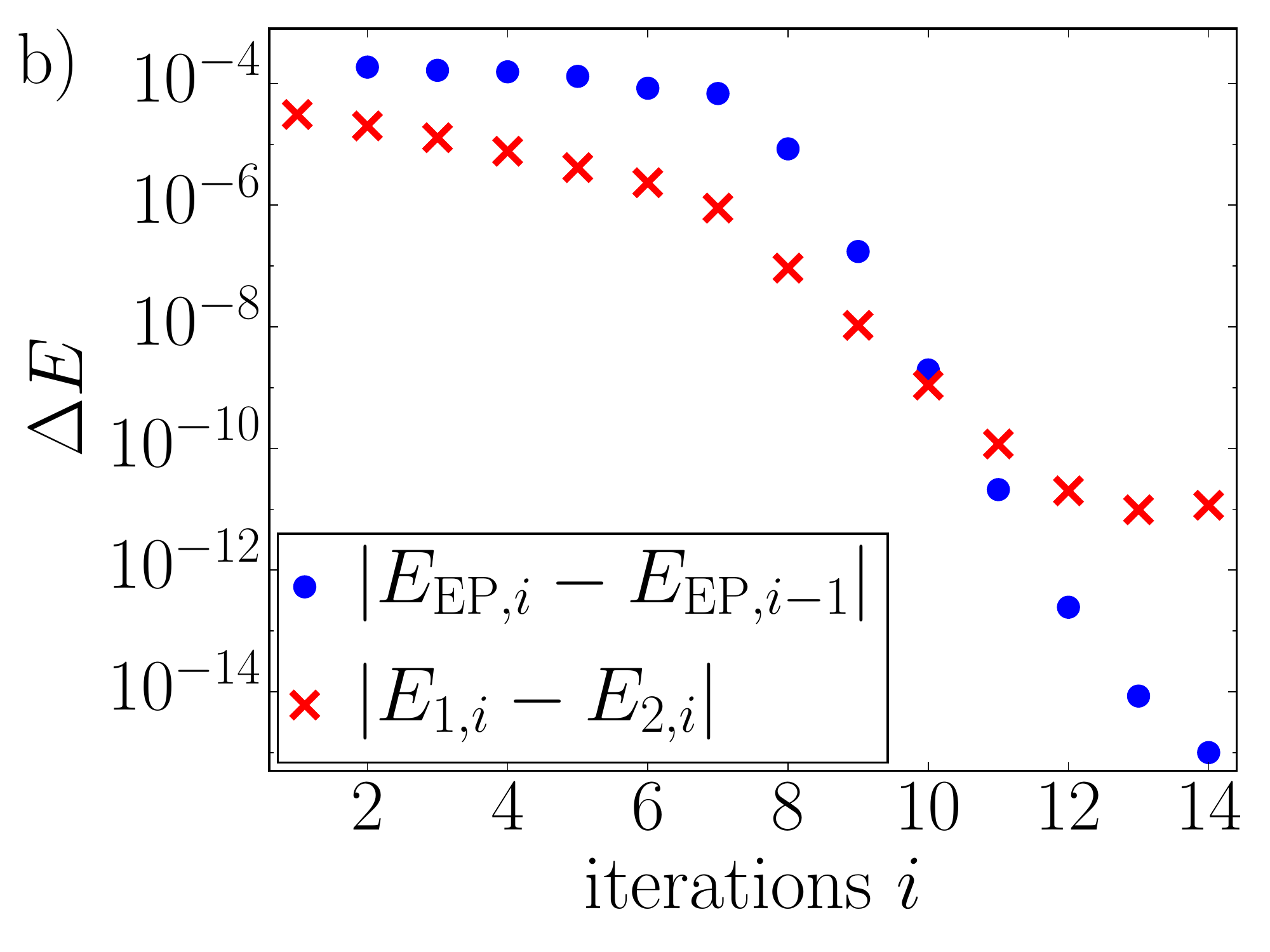}
\caption{a) Real part of the energy of the resonances obtained by solving Eq.~\eqref{generalized} with $\gamma / f = 80$, $m = 0$ and $n\sno{max} = 90$. The levels are labelled by their principal quantum numbers $n$ in the limit $\gamma, f \rightarrow 0$. Two avoided crossings are visible. The right one (marked as a.c.~with a red arrow) between the levels with $n = 10$ and $n = 8$ is taken to obtain a set of initial parameters for the \ac{OM} to search for an \ac{EP} in its vicinity (see text). b) Typical convergence behaviour to an \ac{EP} with the octagon method. Red crosses: modulus of the energy difference $\Delta E$ of the two resonances $E_{1,i}$ and $E_{2,i}$ coalescing to an \ac{EP}, at iteration step $i$. Blue dots: modulus of the difference $\Delta E$ between the estimation $E_{\no{EP},i} = (E_{1,i} + E_{2,i})/2$ for the complex energy of an \ac{EP} and the estimation $E_{\no{EP},i-1}$ of the precedent iteration step.}\label{fig:levels}%
\end{figure}

To find initial parameters for the \ac{OM}, we look at first for avoided crossings between two levels in the term-scheme of the Rydberg system in parallel fields, as level repulsion is associated with the occurrence of \acp{EP}~\cite{heiss2000repulsion}.  For this purpose, the real part of the resonances' energy, which is obtained by solving Eq.~\eqref{generalized}, can be plotted as a function of the field strength $\gamma$ keeping the ratio $\gamma/f$ constant. Fig.~\ref{fig:levels}a shows an excerpt of such spectra for the ratio $\gamma /f = 80$, where a maximum quantum number of $n\sno{max} = 90$ has been used.  An avoided crossing is marked by a red arrow \textcolor{black}{(we could as well choose the other avoided crossing visible in Fig.~\ref{fig:levels}a as long as the correct eigenvalues related to the avoiding states are used to set up the two-dimensional matrix model)}. Here we extract the initial parameters $\gamma_0 = 1.481\times 10^{-3}$, $f_0 = 1.851\times 10^{-5}$ and $\no{Re}(E) = -6.90\times 10^{-3}$ to start our iterative algorithm.

% Initial parameters to find an \ac{EP} are obtained using the field strengths and the real energy of an avoided crossing, like the one that is marked by a red arrow.
% An avoided crossing between the levels of principal quantum number $N = 8$ and $N = 10$ is marked by a red arrow yielding a set of initial parameters $(\gamma_0 = 1.481\times 10^{-3},~f_0 = 1.851\times 10^{-5},~\no{Re}(E) = -6.90\times 10^{-3})$.

% In a different example with a constant ratio of $\gamma / f = 80$, the initial set of parameters $(\gamma_0 = 1.481\times 10^{-3},~f_0 = 1.851\times 10^{-5},~\no{Re}(E) = -6.90\times 10^{-3})$ can be found for the avoided crossing between the levels of principal quantum number $N = 8$ and $N = 10$ (see Fig.~\ref{fig:levels}a).
\begin{figure}[t]
\centering
\includegraphics[width=0.5\columnwidth]{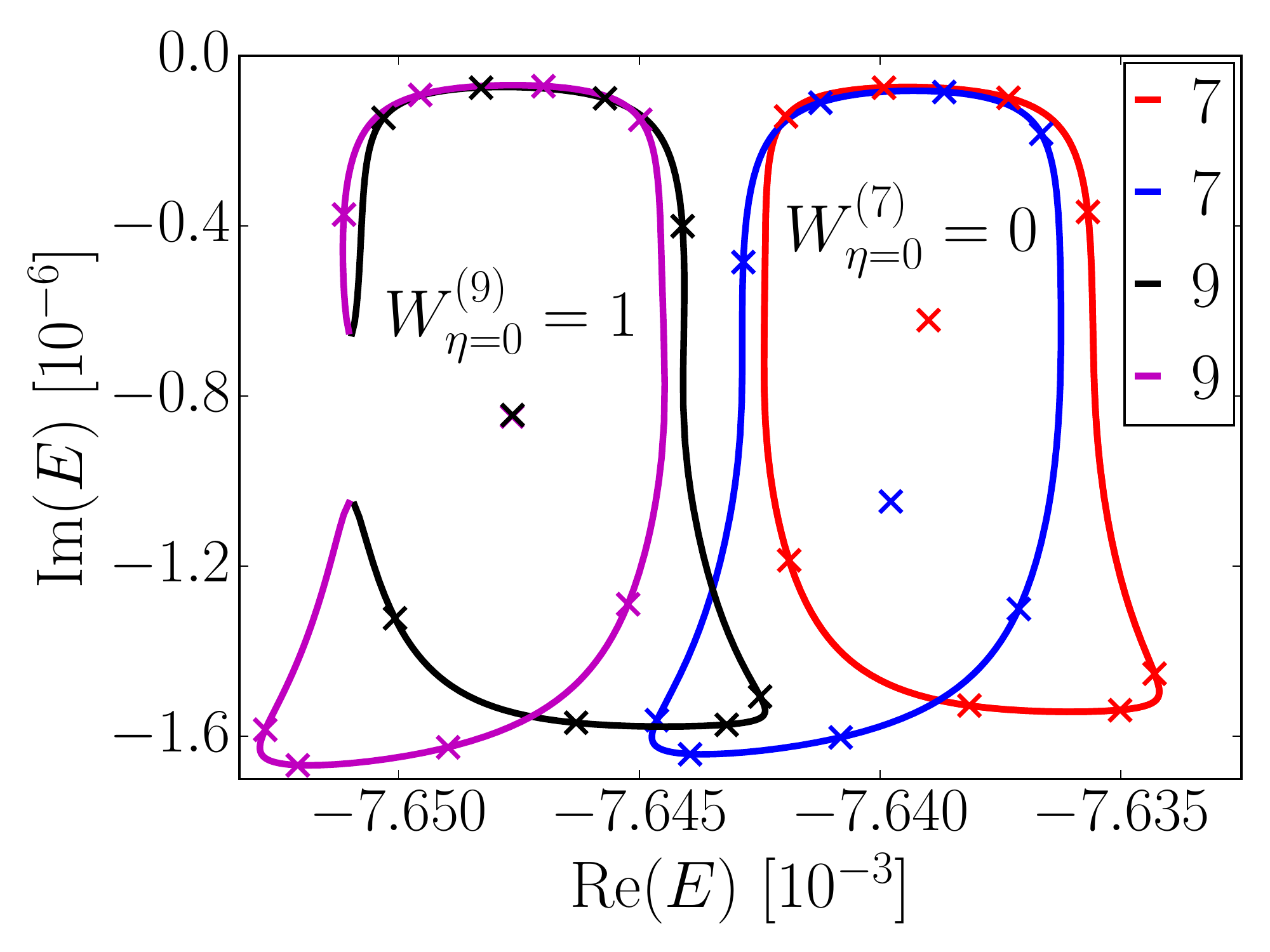}
\caption{Exemplary paths of the resonances in the complex energy plane for the iteration steps $i = 7$ and $i = 9$ of Fig.~\ref{fig:levels}b. The crosses mark the resonances calculated by solving Eq.~\eqref{generalized} at the nine points of the octagon (see Fig.~\ref{fig:oktagon}). Also shown are the resonance energies at the centre-point of the octagons. While for $i=7$ these resonances are still clearly separated, for $i=9$ they coincide within the drawing accuracy. They match the paths that are calculated via Eq.~\eqref{resonances_fit} using the respective coefficients $A$ to $I$ of each iteration step. The path of each resonance is plotted in a different colour. In iteration step seven, each resonance returns to its starting point after a full circle around the octagon. Thus, no \ac{EP} is located within the respective $(\gamma, f)$ region and the winding number according to Eq.~\eqref{winding_number} is $W_{\eta = 0}^{(7)} = 0$. In step nine the resonances show the clear exchange behaviour of an \ac{EP}, which is located close to the centre point of the octagon. The winding number is $W_{\eta = 0}^{(9)} = 1$. In contrast to step seven, the two resonances are almost degenerate at the centre point indicating convergence of the algorithm.}\label{fig:last_steps}%
\end{figure}

Starting from these initial parameters, the \ac{OM} converges within $14$ iteration steps towards the precise position of an \ac{EP} in parameter space
\begin{equation}
\gamma\sno{EP} = 8.598\,633\,574\times 10^{-4},\quad f\sno{EP} = 2.005\,076\,385\times 10^{-5},
\end{equation}
at which the two corresponding resonances become degenerate. The closer the centre-point of the octagon gets to the true position of the \ac{EP}, the smaller becomes the value $|E_{1,i} - E_{2,i}|$ in each iteration step $i$, as is displayed in Fig.~\ref{fig:levels}b (red crosses). For a given matrix representation the degeneracy of the two resonances can only be reached down to an energy difference of $\Delta E = 10^{-10}$. This can be explained by the limited numerical precision when solving Eq.~\eqref{generalized} with the \textit{IRAM} method of the \textit{ARPACK} package~\cite{lehoucq1998arpack}.
% \begin{figure}[t]
% \centering
%   \includegraphics[width=0.49\columnwidth]{D2_plot.pdf}
%   \includegraphics[width=0.49\columnwidth]{gesamtbild.pdf}
%   \caption{a) Typical convergence behaviour of finding an \ac{EP} with the octagon method. Red $\times$: absolute value of the difference at iteration step $i$ between the Energy of the two resonances $E_{1,i}$ and $E_{2,i}$ degenerating to an \ac{EP}. Blue $\bullet$: absolute value of the difference of the estimation $E_{\no{EP},i} = (E_{1,i} + E_{2,i})/2$ for the complex energy of an \ac{EP} relative to the last iteration step. b) Energies of the two \ac{EP} states ($\times$) calculated at the center-point of the octagon in various iteration steps of Fig.~\ref{fig:d2_plot}a. Following equation \eqref{estimation_ep}, the \ac{EP}'s position can by estimated as their mean value ($\circ$), converging to the precise energy of the \ac{EP}.}
%   \label{fig:d2_plot}
% \end{figure}
To estimate the precise complex energy $E\sno{EP}$ of an \ac{EP}, the energy of the two corresponding resonances $E_1$ and $E_2$ can be averaged \textcolor{black}{(by calculating $\gamma\sno{EP}$ and $f\sno{EP}$ for $\eta = 0$ in Eq.~\eqref{2dmatrixentwicklung} and applying the results to Eq.~(\ref{2dmatrixalles}a))}
\begin{equation}
E\sno{EP} = \frac{E_1 + E_2}{2} + {\mathcal O}\left((\gamma_0-\gamma\sno{EP}), (f_0-f\sno{EP})\right).
\label{estimation_ep}
\end{equation}
Being close to  $\gamma\sno{EP}$ and $f\sno{EP}$ within the system, higher order terms can be neglected. While the degeneracy of the corresponding resonances only goes down to an energy difference of $\Delta E = 10^{-10}$, estimation \eqref{estimation_ep} for the complex energy of the \ac{EP} converges with a considerably higher precision, 
\begin{equation}
E\sno{EP} = -7.647\,637\,585\times 10^{-3} - 8.461\,814\,32\times 10^{-7}\no{i},
\end{equation}
during the iteration process (see blue dots in Fig.~\ref{fig:levels}b).\\
To verify that the point of convergence really is an \ac{EP}, we calculate for each iteration step in Fig.~\ref{fig:levels}b the resonances $E_1$ and $E_2$ according to Eq.~\eqref{resonances_fit} using the respective coefficients $A$ to $I$ of the \ac{OM}. As an example we show for an elliptical path circumscribing the octagon in the $(\gamma, f)$ space the resulting paths of the resonances in the complex energy plane (for the two iteration steps $i=7$ and $i = 9$) in Fig.~\ref{fig:last_steps}. While no \ac{EP}-typical exchange behaviour can be seen in iteration step seven, in step nine the two resonances exchange their position after a full circle of the octagon.  This indicates an \ac{EP} located near the centre-point of the octagon, where the two resonances almost become degenerate. To avoid checking the resulting paths for exchange behaviour visually, the winding number can be calculated using Eq.~\eqref{winding_number}. While it is still zero in iteration step seven the winding number becomes one in step nine, and thus the occurrence of an \ac{EP} is verified numerically.

\begin{figure}[t]
\centering
\includegraphics[width=0.5\columnwidth]{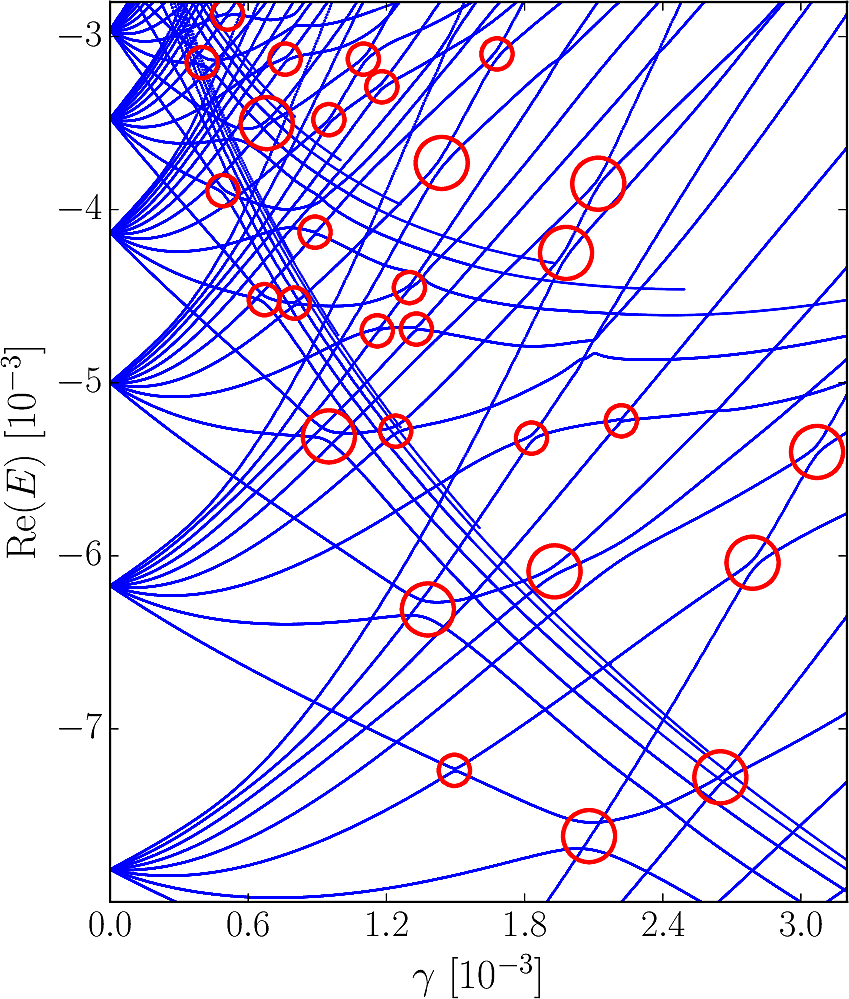}
\caption{Real part of the energy of the resonances obtained by solving Eq.~\eqref{generalized} with maximum quantum number $n\sno{max} = 90$. The strengths $\gamma$ and $f$ are increased simultaneously, keeping the ratio $\gamma / f = 120$ constant. A value of $\gamma = 3\times 10^{-3}$ either corresponds to $B = 705\,\no{T}$ for the hydrogen atom or to $B = 1.8\,\no{T}$ for $\no{Cu}_2\no{O}$ (see \ref{app_hartree}). Avoided crossings are marked by red circles.}\label{fig:avoided_crossing}%
\end{figure}

To find more \acp{EP} for a Rydberg system in parallel fields we generate spectra as in Fig.~\ref{fig:levels}a for a wider range of parameters, see, e.\,g., Fig.~\ref{fig:avoided_crossing}, in which the spectra are calculated with a ratio $\gamma / f = 120$ and a maximum quantum number of $N\sno{max} = 90$. The modulus of the complex dilation parameter $b = |b|\,\exp(\no{i}\alpha)$ has been chosen to be $|b| = \sqrt{32/35}\,\gamma^{-1/6}$ (see Ref.~\cite{muller1994scars}), whereas $\alpha$ has been varied to ensure convergence of the resonances (converged resonances are independent of the rotation angle $\alpha$). The resulting spectra cover a range of $\gamma$ up to $\gamma = 3\times 10^{-3}$, which corresponds to $B = 705\,\no{T}$ for the hydrogen atom (still far away from being realised experimentally yet) or to $B = 1.8\,\no{T}$ for $\no{Cu}_2\no{O}$, which is fully accessible in a possible experimental realisation (see, e.\,g., Ref.~\cite{assmann}). Figure \ref{fig:avoided_crossing} shows a high number of avoided crossings (most of them are marked by red circles). From each of them an initial set of parameters for the \ac{OM} can be extracted to start the iterative search for an \ac{EP}.

% In contrast to the hydrogen atom, where $\gamma = 3\times 10^{-3}$ corresponds to $B = 705\,\no{T}$ (which is still far away from being realized experimentally yet), the according value for $\no{Cu}_2\no{O}$ is with $B = 3.3\,\no{T}$ fully accessible in a possible experimental realization (see, e.\,g., Ref.~\cite{assmann}).

\begin{figure}[t]
\centering
\includegraphics[width=0.65\columnwidth]{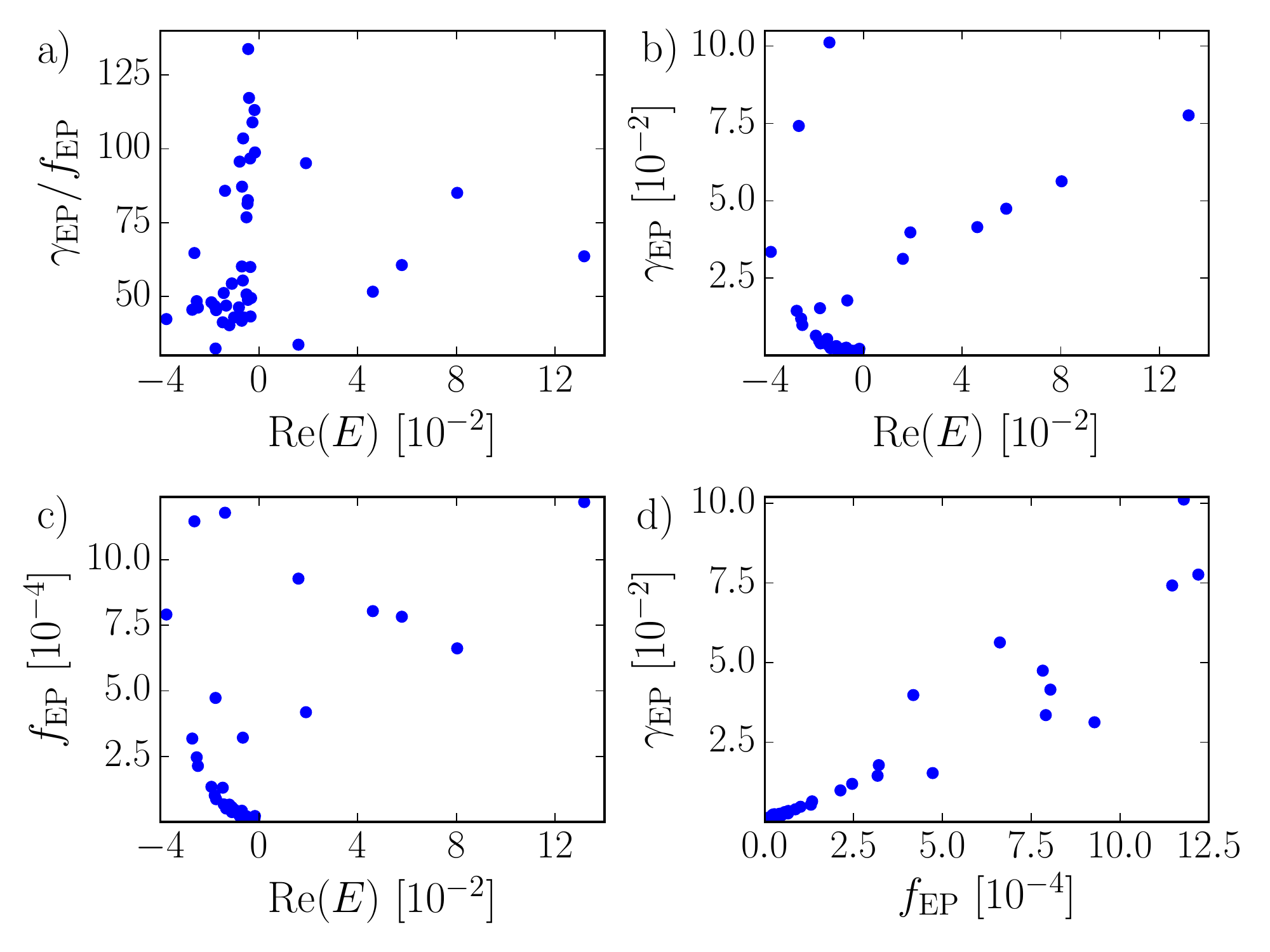}
\caption{\acp{EP} found in the system described by Eq.~\eqref{hamiltonian_hartree} using the \ac{OM}. The \acp{EP} are presented in the material-independent description of appropriate units. These results can be converted either to values for the hydrogen atom or to values for $\no{Cu}_2\no{O}$ using the relations in \ref{app_hartree}.}\label{fig:ep_data}%
\end{figure}

With the \ac{OM} several \acp{EP} could be found (see Fig.~\ref{fig:ep_data}) in the system described by Eq.~\eqref{hamiltonian_hartree}. Using the proper constants $F_0$ and $B_0$ as well as the correct energy scaling (see \ref{app_hartree}), the results can be converted to SI-units either for the hydrogen atom or for excitons in $\no{Cu}_2\no{O}$. A large portion of the \acp{EP} shown in Fig.~\ref{fig:ep_data} could be found in a region with $\no{Re}(E)<0$ (the zero-point of the $\no{Re}(E)$-axis is defined by the threshold between bound states and continuum states in the absence of external fields) and relatively small external fields up to ($\gamma\sno{EP}\approx 10^{-2}$, $f\sno{EP} \approx 10^{-4}$). This corresponds to ($\gamma\sno{EP}\approx 2.35\times 10^{3}\,\no{T}$, $f\sno{EP}\approx 5.14\times 10^{5}\,\no{V/cm}$) for the hydrogen atom or to ($\gamma\sno{EP}\approx 6.03\,\no{T}$, $f\sno{EP}\approx 1.76\times 10^{2}\,\no{V/cm}$) for $\no{Cu}_2\no{O}$. In Fig.~\ref{fig:ep_data}d one can see that the \acp{EP} found are approximately located along a straight line with $\gamma\sno{EP} / f\sno{EP} = 50$ in this low-field regime (the precise values for $\gamma\sno{EP}/f\sno{EP}$ are displayed in Fig.~\ref{fig:ep_data}a, where the accumulation near $50$ is less obvious). At higher fields a significant deviation from this behaviour is found. Even though the \acp{EP} seem to form two branches if only the magnetic field is considered, see Fig.~\ref{fig:ep_data}b, in this high-field regime such a behaviour cannot be observed for the electric field in Fig.~\ref{fig:ep_data}c. However, both figures show the same behaviour for small fields: the higher the strength of the external fields, the lower are the energies of the resonances of the Rydberg system which interact to form an \ac{EP}. When reaching the high-field regime, this behaviour changes and \acp{EP} with $\no{Re}(E)>0$ can be found, too. It is important to note that Fig.~\ref{fig:ep_data} does not show all existing \acp{EP} of a Rydberg system in parallel electric and magnetic fields but just those found within the scope of our work. Therefore, more \acp{EP} could be found, which possibly would change the appearance of Fig.~\ref{fig:ep_data} especially in the high-field regime.

In table~\ref{tab:found_ep} we present the parameters of selected \acp{EP} to compare the significantly differing physical values for the hydrogen atom and $\no{Cu}_2\no{O}$. With regard to the external field strengths of the \acp{EP}, the fields for $\no{Cu}_2\no{O}$ are about two orders of magnitude smaller than those for the hydrogen atom due to effective masses and the relative dielectric constant of the solid (see \ref{app_hartree}). In contrast to the hydrogen atom, the external fields to detect \acp{EP} in $\no{Cu}_2\no{O}$ can be realised in an experiment (see, e.\,g., Ref.~\cite{assmann} with $B$ up to $7\,\no{T}$). Hence, we propose the measurement of photoabsorption spectra of the states associated with an \ac{EP} in $\no{Cu}_2\no{O}$. Close to the \ac{EP} these spectra can be analysed by means of harmonic inversion~\cite{cartarius2007exceptional} to detect the exchange behaviour of resonances and to use the \ac{OM} to locate the precise position of an \ac{EP} in the real system of $\no{Cu}_2\no{O}$. Extending this method according to Ref.~\cite{fuchs2014harmonic}, even an investigation directly at the \ac{EP} would be possible.

\begin{table}
\caption{Parameters of selected \acp{EP} of the system described by Eq.~\eqref{hamiltonian_hartree}. Note that the energies of $\no{Cu}_2\no{O}$ are given without the offset caused by the band gap energy of $E\sno{gap} = 2.17208\,\no{eV}$~\cite{kazimierczuk2014giant}.}
\label{tab:found_ep}
\centering
\begin{tabular}{llll|llll}
\toprule
\multicolumn{4}{c}{Hydrogen atom} & \multicolumn{4}{c}{$\no{Cu}_2\no{O}$}\\
% \cline{1-2}
% \cline{4-5}
$B$ [$\no{T}$] & $F$ [$\frac{\no{V}}{\no{cm}}$] & $E\sno{r}$ [$\no{eV}$] &$E\sno{i}$ [$\no{meV}$] & $B$ [T] & $F$ [$\frac{\no{V}}{\no{cm}}$] &$E\sno{r}$ [$\no{meV}$] & $E\sno{i}$ [$\upmu\no{eV}$]\\
\midrule
% 229.64 & 120250  & $-0.1904$ & $-0.6209$ & 0.590 1.096& 41.16 & $-1.286$ & $-0.619$ \\
% 561.26 & 140870  & $-0.1866$ & $-0.2564$ & 1.441 2.678& 48.22 & $-1.261$ & $-1.732$ \\
% 799.69 & 341940  & $-0.3886$ & $-2.072$ & 2.053 3.815& 117.0 & $-2.625$ & $-14.00$ \\
% 1261.3 & 668930  & $-0.3996$ & $-0.5002$ & 3.238 6.017& 229.0 & $-2.699$ & $-3.379$ \\
% 1506.7 & 686310  & $-0.5245$ & $-4.402$ & 3.868 7.188& 234.9 & $-3.544$ & $-29.74$ \\
% 2316.3 & 1096200 & $-0.6733$ & $-0.5999$ & 5.946 11.05& 375.2 & $-4.549$ & $-4.054$ \\
% 3595.7 & 2430880 & $-0.4788$ & $-12.03$ & 9.231 17.15& 832.0 & $-3.234$ & $-81.25$ \\
229.64 & \phantom{0}120250  & $-0.1904$ & $-0.6209$ & 0.590& 41.16 & $-1.286$ & $-0.419$ \\
561.26 & \phantom{0}140870  & $-0.1866$ & $-0.2564$ & 1.441& 48.22 & $-1.261$ & $-1.732$ \\
799.69 & \phantom{0}341940  & $-0.3886$ & $-2.072$ & 2.053& 117.0 & $-2.625$ & $-14.00$ \\
1261.3 & \phantom{0}668930  & $-0.3996$ & $-0.5002$ & 3.238& 229.0 & $-2.699$ & $-3.379$ \\
1506.7 & \phantom{0}686310  & $-0.5245$ & $-4.402$ & 3.868& 234.9 & $-3.544$ & $-29.74$ \\
2316.3 & 1096200 & $-0.6733$ & $-0.5999$ & 5.946& 375.2 & $-4.549$ & $-4.054$ \\
3595.7 & 2430880 & $-0.4788$ & $-12.03$ & 9.231& 832.0 & $-3.234$ & $-81.25$ \\
\bottomrule
\end{tabular}
\end{table}

\begin{figure}[t]
\centering
\includegraphics[width=0.65\columnwidth]{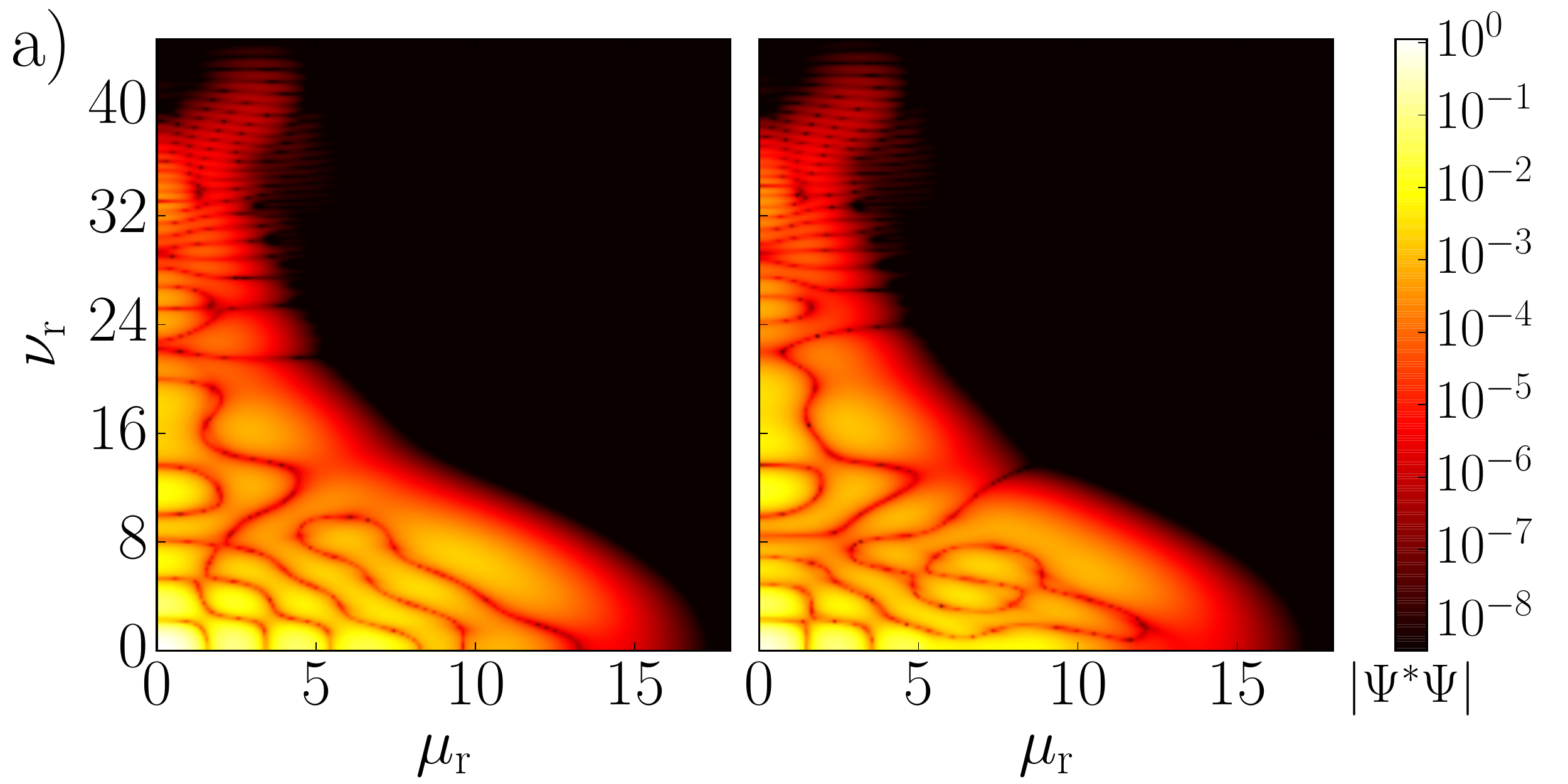}
% \caption{Visualization of the probability density $|\Psi^*\Psi|$ for the two states belonging to the \ac{EP} at $(\gamma = 2.387\,819\times 10^{-3}, f = 2.739\,422\times 10^{-5})$ with the energy $E = -0.685\,886\times 10^{-2} - 0.942\,211 \times 10^{-5}\no{i}$ in the $\mu\sno{r}$- and $\nu\sno{r}$-plane of semiparabolic coordinates. For further information see text.}\label{fig:wavefunctions_at_ep_1}%
\includegraphics[width=0.52\columnwidth]{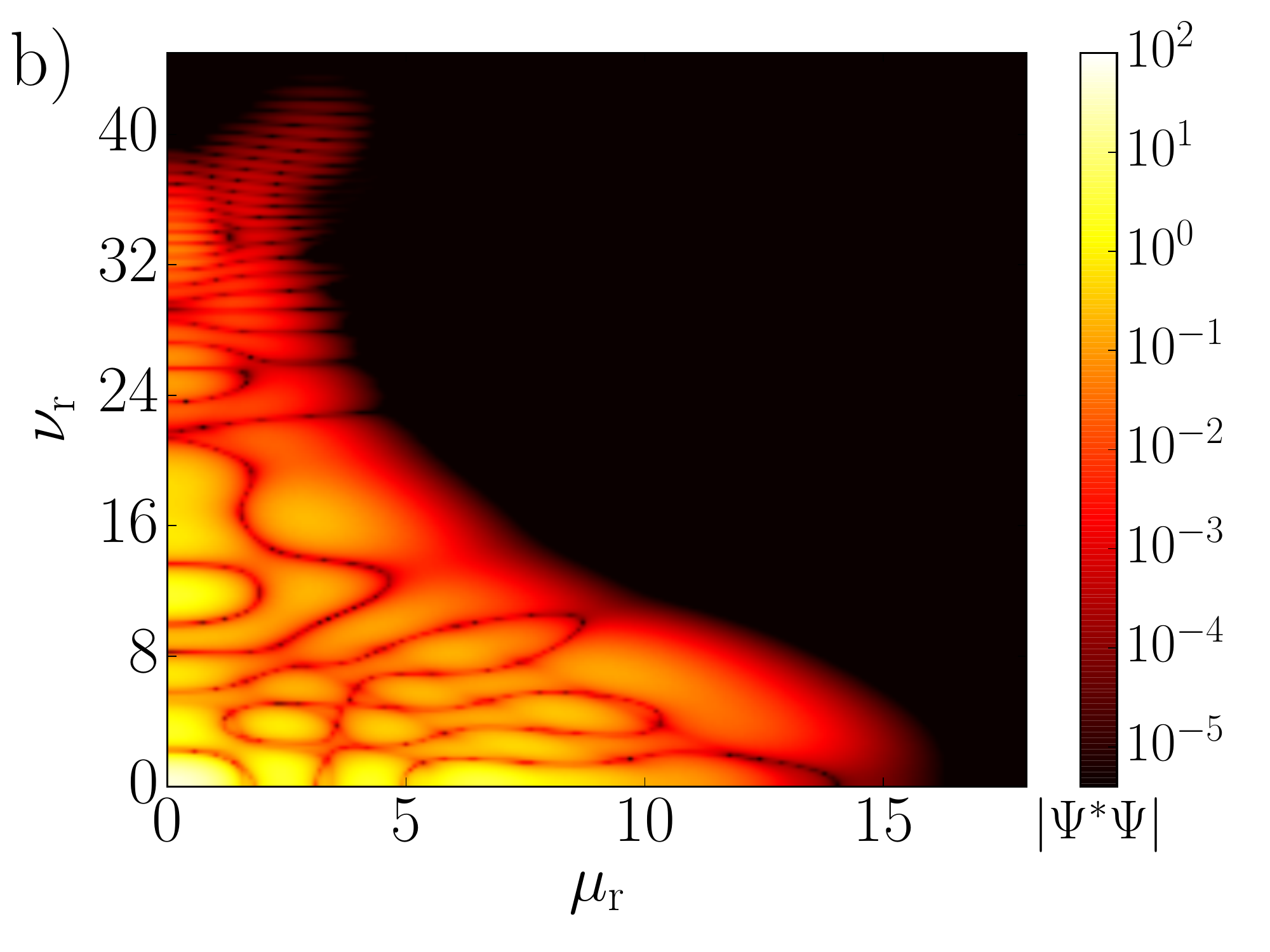}
% \caption{Estimation for the probability density $|\Psi^*\Psi|$ of the degenerated state at the \ac{EP} of Fig.~\ref{fig:wavefunctions}a. It is calculated according to equation \eqref{wave_functions_semiparabolic} based on the estimation $\left|v\sno{EP}\right>\sno{guess}$ for the degenerated eigenvector (see Eq.~\eqref{state_degeneration}).}\label{fig:wavefunctions_at_ep_mean_eigenvalues}%
\caption{a) Visualisation of the probability density $|\Psi^*\Psi|$ for the two states belonging to the \ac{EP} at $(\gamma = 2.387\,819\times 10^{-3}, f = 2.739\,422\times 10^{-5})$ with the energy $E = -0.685\,886\times 10^{-2} - 0.942\,211 \times 10^{-5}\no{i}$ in the $(\mu\sno{r}, \nu\sno{r})$-plane of semiparabolic coordinates. For further details see text. b) Estimation for the probability density $|\Psi^*\Psi|$ of the degenerate state at the \ac{EP} of Fig.~\ref{fig:wavefunctions}a. The distribution is calculated according to Eq.~\eqref{wave_functions_semiparabolic} based on the estimation $\left|v\sno{EP}\right>\sno{guess}$ for the degenerated eigenvector (see Eq.~\eqref{state_degeneration}).}
\label{fig:wavefunctions}%
\end{figure}

% \subsection{Wave-functions at an \ac{EP}}
\label{sec:wavefunctions}
To find the precise position of an \ac{EP} in parameter space as well as its exact energy only the eigenvalues of Eq.~\eqref{generalized} are needed. Additionally, using the eigenvectors according to Eq.~\eqref{wave_functions_semiparabolic}, the associated states can be calculated. A visualisation of the probability density $|\Psi^*\Psi|$ for the two states belonging to the \ac{EP} at $(\gamma = 2.387\,819\times 10^{-3}, f = 2.739\,422\times 10^{-5})$ with the energy $E = -6.85\,886\times 10^{-3} - 9.42\,211 \times 10^{-6}\no{i}$ is presented in Fig.~\ref{fig:wavefunctions}a, which is calculated with $n\sno{max} = 90$ and $|b|=2.6$ in semiparabolic coordinates. Due to the cylindrical symmetry of a state with $m=0$ in parallel electric and magnetic fields, only the first quadrant of the $(\mu\sno{r}, \nu\sno{r})$-plane is needed to cover the full position space. The angle $\varphi$ of Eq.~\eqref{semiparabolic} can be chosen arbitrarily and we use $\varphi = 0$. To obtain a real-valued probability density, the absolute value of $\Psi^*\Psi$ has to be taken \cite{moiseyev2011non}. The reason for this is the complex rotation, where only the intrinsic complex parts of $\Psi$ are conjugated in $\Psi^*$ (see Eq.~\eqref{wave_functions_semiparabolic}) and thus $\Psi^*\Psi$ stays a complex quantity.
According to Ref.~\cite{kato1976perturbation} not only the eigenvalues, but also the eigenvectors coalesce at an \ac{EP}. The visualisation of the two associated states should therefore look the same, which cannot be verified in Fig.~\ref{fig:wavefunctions}a. Although the two probability densities have the same extension in the space of semiparabolic coordinates they differ by their nodal line patterns\footnote{The use of the term \emph{nodal lines} must be considered with caution here because at these lines the probability density is just significantly smaller compared to the surrounding area \textendash~it does not go to zero, which would be the common definition for a nodal line.}.

If the nodal patterns of the wave functions of the two states are studied for a circle around the \ac{EP} they show the same exchange behaviour as the eigenvalues. Following a full circle in parameter space around the \ac{EP} in small steps and calculating and sorting the plots of the probability densities for each step, we obtain a transformation of the nodal patterns into each other~\cite{feldmaier2015untersuchung}. Note that we only use the modulus $|\Psi^*\Psi|$ and ignore the phase of the wave function, and thus do not uncover the typical phase change $[\Psi_1, \Psi_2] \rightarrow [\Psi_2, -\Psi_1]$ of the eigenvectors. In principle this can be recovered by tracing the nodal lines during the circle \cite{dembowski2001experimental}.

The reason for the difference in the nodal patterns of the two states in Fig~\ref{fig:wavefunctions}a, which have been calculated directly at the precise position of the \ac{EP} can be found by looking at Fig.~\ref{fig:levels}b, where the degeneracy of the eigenvalues (red $\times$) can only be reached down to an energy difference of $\Delta E = 10^{-10}$ due to numerical limitations. The same limitations prevent the eigenvectors from degenerating at the \ac{EP}. The accuracy of calculating the exact energy at an \ac{EP} could be raised by taking the mean value of the two related resonances (see Eq.~\eqref{estimation_ep} and blue dots in Fig.~\ref{fig:levels}b). Extending this approach to the eigenvectors, the guessed (degenerate) eigenvector $\left|v\sno{EP}\right>\sno{guess}$ for the two resonances at the \ac{EP} can be found by
\begin{equation}
\left|v\sno{EP}\right>\sno{guess} = \frac{\left|v_1\right> + \left|v_2\right>}{2},
\label{state_degeneration}
\end{equation}
where $\left|v_1\right>$ and $\left|v_2\right>$ are the (numerically) not completely degenerate eigenvectors that were used to obtain Fig.~\ref{fig:wavefunctions}a. The naive expectation is that it should be possible to calculate the wave function at an \ac{EP} by normalising $\left|v\sno{EP}\right>\sno{guess}$. However, this is in principle impossible since the physically correct inner product $\int \Psi^*\Psi\,\no{d}^3r$ of a wave function at the \ac{EP} vanishes \cite{moiseyev2011non}. That is, close to the \ac{EP} the normalised wave function diverges. This can be observed in Fig.~\ref{fig:wavefunctions}b. Note that the scale for the modulus of the wave function is two orders of magnitude higher compared to Fig.~\ref{fig:wavefunctions}a. Consequently, the level of divergence \textendash~or in other words the smallness of $\int \Psi^*\Psi\,\no{d}^3r$ \textendash~is a further indicator of how close $\left|v\sno{EP}\right>\sno{guess}$ is to the exact position of the \ac{EP}.

% By normalising $\left|v\sno{EP}\right>\sno{guess}$, the probability distribution of the degenerated state at the \ac{EP} can be calculated (see Fig.~\ref{fig:wavefunctions}b).
% % \begin{figure}[t]
% % \centering
% % \includegraphics[width=0.5\columnwidth]{wavefunctions_at_ep_mean_eigenvalues.pdf}
% % \caption{Estimation for the probability density $|\Psi^*\Psi|$ of the degenerated state at the \ac{EP} of Fig.~\ref{fig:wavefunctions}a. It has been calculated according to equation \eqref{wave_functions_semiparabolic} based on the estimation $\left|v\sno{EP}\right>\sno{guess}$ for the degenerated eigenvector (see Eq.~\eqref{state_degeneration}).}\label{fig:wavefunctions_at_ep_mean_eigenvalues}%
% % \end{figure}
% Note that the scale in Fig.~\ref{fig:wavefunctions}b is approximately two orders of magnitude higher compared to Fig.~\ref{fig:wavefunctions}a. At an \ac{EP} the eigenvectors of the associated states cannot be normalised that means close to the \ac{EP} the normalisation results in divergences. Hence, the level of divergence is another indicator of how close the calculation reaches the exact position of the \ac{EP}.

\section{Conclusion and outlook}
\label{sec:conclusion-outlook}
Based on a work on \acp{EP} in the hydrogen atom in \emph{crossed} electric and magnetic fields~\cite{cartarius2007exceptional}, we started to investigate the appearance of \acp{EP} in a generalised and more symmetric model: the hydrogen-like system in \emph{parallel} electric and magnetic fields. Using appropriate units (see \ref{app_hartree}), our results hold either for the hydrogen atom or for $\no{Cu}_2\no{O}$ (for which a hydrogen-like spectrum of excitons up to a principal quantum number of $n =25$ has been found recently~\cite{kazimierczuk2014giant}). The initial points to start an iterative search for the precise position of an \ac{EP} in parameter space are given by the parameters of an avoided crossing within the spectra because level repulsion is associated with the occurrence of \acp{EP}~\cite{heiss2000repulsion}. Since the common 3-point method \cite{uzdin2010finding} of finding \acp{EP} in open quantum systems has to start its iterative search close to an \ac{EP}'s actual position, it does in general not show convergence with the initial parameters of an avoided crossing. Using a two-dimensional matrix model to describe the two states forming an \ac{EP} in its local vicinity, we have developed the \ac{OM}, an iterative method that converges to the precise position of an \ac{EP} in parameter space. Performing the time-consuming calculation of the resonances at just nine points in parameter space (eight at the corners of an octagon and one at its centre-point), the \ac{OM} provides a very fast and stable method which only needs the initial parameters of an avoided crossing to converge to the precise position of the related \ac{EP} in parameter space, and which yields its precise complex energy by taking the mean energy of the two associated resonances. Using the \ac{OM}, the paths of the associated resonances in the complex energy plane during a circle around the \ac{EP} in small steps in parameter space can be revealed without further quantum-mechanical (and time-consuming) calculations. The exchange behaviour of these paths can be tested numerically by calculating a winding number which verifies the existence of an \ac{EP}. Using the \ac{OM}, we were able to find a variety of \acp{EP} within the open quantum system of an exciton in parallel electric and magnetic fields. In contrast to the hydrogen atom, a variety of the \acp{EP} that we predict for $\no{Cu}_2\no{O}$ are in a regime below $B = 10\,\no{T}$, which is experimentally accessible. Hence, $\no{Cu}_2\no{O}$ provides an appropriate system to verify our theoretical predictions experimentally. A possible experimental realisation would be the measurement of photoabsorption spectra of the states associated with an \ac{EP} in $\no{Cu}_2\no{O}$ and analysing them by means of harmonic inversion~\cite{cartarius2007exceptional}. Even an extension according to Ref.~\cite{fuchs2014harmonic} to investigate spectra directly at the \ac{EP} would be possible.

In the second part of our work we visualised the coalescence of the two wave functions belonging to the resonances which become degenerate at the \ac{EP}. Calculating the probability densities in semiparabolic coordinates at the precise position of the \ac{EP}, the exact coalescence could not be confirmed for numerical reasons as the two states differ by their nodal line patterns (these patterns show the exchange behaviour expected for the two resonances since they transform into each other while completely encircling the \ac{EP} in small steps~\cite{feldmaier2015untersuchung}). Motivated by a mean-value-calculation for the associated complex resonances to determine the precise energy of an \ac{EP}, we averaged the numerically different eigenvectors of the two resonances directly at the \ac{EP} and obtained one degenerate eigenvector, of which we calculated the probability density.

In an experiment the realisation of an \ac{EP} in hydrogen-like resonance spectra seems to be very promising for excitons in semiconductor devices \cite{assmann, thewes2015observation}. In this context our work gives clear evidence for the appearance of \acp{EP}. However, extensions to some simplifications should be investigated. From a theoretical point of view some work still remains to be done to extend the hydrogen-like model for excitons in $\no{Cu}_2\no{O}$ to a more realistic model. One should, e.\,g., include a more complex band structure with more than one valence band. Another possibility would be to consider the exchange and correlation effects according to Ref.~\cite{roessler2009solid} or to include exciton-phonon interaction. Furthermore, we plan to perform calculations for $\no{Cu}_2\no{O}$ in crossed electric and magnetic fields, as it was done for the hydrogen atom in, e.\,g., Refs.~\cite{cartarius2007exceptional,schweiner2015classical,Diss30,Diss31,PRA75_1,PRA75_3,Diss14}. For this case, the Hamiltonian~\eqref{ham_ex} for $\no{Cu}_2\no{O}$ cannot be brought to the same form as the Hamiltonian~\eqref{ham_hyd} for the hydrogen atom by using an appropriate scaling. Consequently, the results for the hydrogen atom cannot be converted to the results for $\no{Cu}_2\no{O}$ in a simple way and separate calculations need to be done.

% \ack

\appendix

\section{Choice of appropriate units}
\label{app_hartree}
% A set of appropriate units to describe the hydrogen atom are Hartree-Units, which are introduced by setting $\hbar = 4\pi\,\varepsilon_0 = m\sno{e} = e = 1$, what makes these quantities dimensionless (the unit a.u. can be written formally here). 

The use of appropriate units within the course of the present work has the advantage that calculations can be carried out without material-dependent parameters (see Eq.~\eqref{hamiltonian_hartree}). These parameters are absorbed in the scaling constants $F_0$ of the electric and $B_0$ of the magnetic field. The numerical value of these constants depends on the system considered; either a hydrogen atom or a Rydberg exciton in $\no{Cu}_2\no{O}$ (see Tab.~\ref{tab:hartree}). \textcolor{black}{The difference in these constants can be explained on one hand by the reduced mass $\mu$. The mass of the proton in the hydrogen atom can be taken as approximately infinite compared to the mass of the electron. Hence, $\mu \approx m_0$ here (with $m_0$ being the mass of an electron in free space), whereas for Rydberg excitons in $\no{Cu}_2\no{O}$ the reduced mass amounts to $\mu = 0.38m_0$~\cite{sasaki1973magneto}. On the other hand, the Coulomb interaction in $\no{Cu}_2\no{O}$ is screened by the dielectric constant $\varepsilon\sno{r} = 7.50$~\cite{roessler2009solid}. In the case of a hydrogen atom there is no such screening, and thus $\varepsilon\sno{r} = 1$.}

% As explained in Sec.~\ref{sec:methods}, the two equations \eqref{ham_hyd} and \eqref{ham_ex} can be written in the same form~\eqref{hamiltonian_hartree}.
% by introducing a scaling factor $\zeta$ according to equation \eqref{ham_ex} into the magnetic flux density $B_0$ (see Tab.~\ref{tab:hartree})
% \begin{equation}
% \zeta = \sqrt{\frac{m\sno{h}^3 + m\sno{e}^3}{\left(m\sno{h} + m\sno{e}\right)^3}} \approx \begin{cases}  0.538 &\text{for}~\no{Cu}_2\no{O},\\  1 &\text{for H atom}. \end{cases}
% \end{equation}
% For Rydberg excitons in $\no{Cu}_2\no{O}$ this factor can be calculated using an effective electron mass of $m\sno{e} = 0.99m_0$ and an effective hole mass of $m\sno{h} = 0.62m_0$~\cite{sasaki1973magneto}(with $m_0$ being the mass of an electron in free space). In case of a hydrogen atom the scaling factor yields $\zeta = 1$, as $m\sno{h}$ now corresponds to the proton mass, which can be taken as infinite compared to the mass $m\sno{e}=m_0$ of the electron.
\textcolor{black}{Having done a calculation in appropriate units the quantities in table~\ref{tab:hartree} can be used to convert the result to either hold for a hydrogen atom or for Rydberg excitons in $\no{Cu}_2\no{O}$. As an example: a magnetic flux density of $\gamma = 1$ in Eq.~\eqref{hamiltonian_hartree} corresponds either to $B_0= 2.35\times 10^{5}\,\no{T}$ for the hydrogen atom or to $B_0 = 603\,\no{T}$ for $\no{Cu}_2\no{O}$. 
}

\begin{table}[h]
\centering
\caption{Appropriate units converted to SI-units. The values for the hydrogen atom are taken from~\cite{mohr2005codata}. Here we assume, that the mass of the proton is infinite as compared to the mass of the electron. In this case the appropriate units correspond to \emph{Hartree units}. The values for $\no{Cu}_2\no{O}$ are calculated according to the given formulas using the reduced mass $\mu = 0.38m_0$~\cite{sasaki1973magneto} and $\varepsilon\sno{r} = 7.50$~\cite{roessler2009solid}.}
\label{tab:hartree}
\begin{tabular}{llll}
\toprule
quantity & appropriate unit & value H atom & value $\no{Cu}_2\no{O}$\\
\midrule
energy & $E\sno{h} = \mu\,e^4 / (4\pi\,\varepsilon_0\,\varepsilon\sno{r}\,\hbar)^2$ & $4.359\,744\times 10^{-18}\,\no{J}$ & $2.945\times 10^{-20}\,\no{J}$\\
%% Time & $\hbar / E\sno{h}$ & $2.418\,884 \times 10^{-17}\,\no{s}$ & $3.581\times 10 ^{-15}\,\no{s}$\\
length & $a_0 = 4\pi\,\varepsilon_0\,\varepsilon_r\,\hbar^2 / (\mu\,e^2)$ & $0.529\,177\times 10^{-10}\,\no{m}$ & $1.044\times 10^{-9}\,\no{m}$\\
%% Mass & $m\sno{e}$ & $9.109\,382\times 10^{-31}\,\no{kg}$ & $3.462\times 10^{-31}\,\no{kg}$\\
%% Charge & $e$ & $1.602\,176\times 10^{-19}\,\no{C}$ & $1.602\times 10^{-19}\,\no{C}$\\
%% Momentum & $\hbar/a_0$ & $1.992\,852\times 10^{-24}\,\frac{\no{kg}\,\no{m}}{\no{s}}$ & $1.010\times10^{-25}\,\frac{\no{kg\,m}}{\no{s}}$\\
%% Angular momentum & $\hbar = h/(2\pi)$ & $1.054\,572\times 10^{-34}\,\no{Js}$ & $1.055\times 10^{-34}\,\no{Js}$\\
el. field strength & $F_0 = E\sno{h} / (e\,a_0)$ & $5.142\,206\times 10^{11}\,\frac{\no{V}}{\no{m}}$ & $1.760\times 10^{8}\,\frac{\no{V}}{\no{m}}$\\
magn. flux density & $B_0 = \hbar /(e\,a_0^2)$ & $2.350\,517\times 10^{5}\,\no{T}$ & \textcolor{black}{$6.034\times 10 ^{2}\,\no{T}$}\\
\bottomrule
\end{tabular}
\end{table}

\section{Estimating the position of an \ac{EP}}
\label{app_solution}
In Sec.~\ref{sec:octagon_method} a solution of Eq.~\eqref{null} is needed to estimate the position $(\gamma\sno{EP}, f\sno{EP})$ of an \ac{EP}. Since the coefficients $D$ to $I$ are complex, both the real and the imaginary part of the polynomial in Eq.~\eqref{null} must be set to zero
\begin{subequations}
\begin{align}
0 &= \no{Re}(D) + \no{Re}(E)\,x + \no{Re}(F)\,y + \no{Re}(G)\,x^2 + \no{Re}(H)\,x\,y + \no{Re}(I)\,y^2\label{oktagon_null_eins},\\
0 &= \no{Im}(D) + \no{Im}(E)\,x + \no{Im}(F)\,y + \no{Im}(G)\,x^2 + \no{Im}(H)\,x\,y + \no{Im}(I)\,y^2.
\label{oktagon_null_zwei}
\end{align}
\label{oktagon_null}%
\end{subequations}
Dividing Eq.~\eqref{oktagon_null_eins} by $\no{Re}(I)$ and Eq.~\eqref{oktagon_null_zwei} by $\no{Im}(I)$ and subtracting the resulting equations cancels the $y^2$-term and yields 
\begin{equation}
y(x) = -\frac{W(D, I) + W(E, I)\,x + W(G, I)\,x^2}{W(F, I) + W(H, I)\,x},
\label{ypsilonvonx}
\end{equation}
where we introduce the abbreviation
\begin{equation}
W(U,V) \equiv U\sno{i}\,V\sno{r} - U\sno{r}\,V\sno{i}
\end{equation}
with $U\sno{i} \equiv \no{Im}(U)$, $U\sno{r} \equiv \no{Re}(U)$ and $U, V \in \{D, E, F, G, H, I\}$. Inserting Eq.~\eqref{ypsilonvonx} in Eq.~\eqref{oktagon_null_eins} for $W(F, I) + W(H, I)\,x \ne 0$ yields a fourth-order polynomial of which the roots need to be found:
\begin{align}
\begin{aligned}
0 =& \phantom{\left\}\right.}f_1\, W(F, I) + I\sno{r}\,W(D, I)^2\\
+& \left\{ f_1\,W(H, I) + f_2\,W(F, I) + 2I\sno{r}\,W(D, I)\,W(E, I)\right\}\,x\\
+& \left\{ f_2\,W(H, I) + f_3\,W(F, I) + 2I\sno{r}\,W(D, I)\,W(G, I) + I\sno{r}\,W(E, I)^2\right \}\,x^2\\
+& \left\{ f_3\,W(H, I) + \left[ 2I\sno{r}\,W(E, I) - H\sno{i}\,W(F, I)\right]\,W(G, I)\right \}\,x^3\\
+& \left\{ \left[H\sno{i}\,W(H, I) + I\sno{r}\,W(G, I)\right]\,W(G, I)\right \}\,x^4.
\end{aligned}
\label{polynomvier}
\end{align}
Here, further abbreviations are used
\begin{subequations}
\begin{align}
f_1 &= D\sno{r} - F\sno{r}\,W(D, I),\\
f_2 &= E\sno{r} - F\sno{r}\,W(E, I) - H\sno{i}\,W(D, I),\\
f_3 &= G\sno{r} - F\sno{r}\,W(G, I) - H\sno{i}\,W(E, I).
\end{align}
\label{abkuerzungen}%
\end{subequations}
Having found a root $x$ of Eq.~\eqref{polynomvier}, the corresponding $y$ can be calculated using Eq.~\eqref{ypsilonvonx}.

\begin{figure}[t]
\centering
\includegraphics[width=0.5\columnwidth]{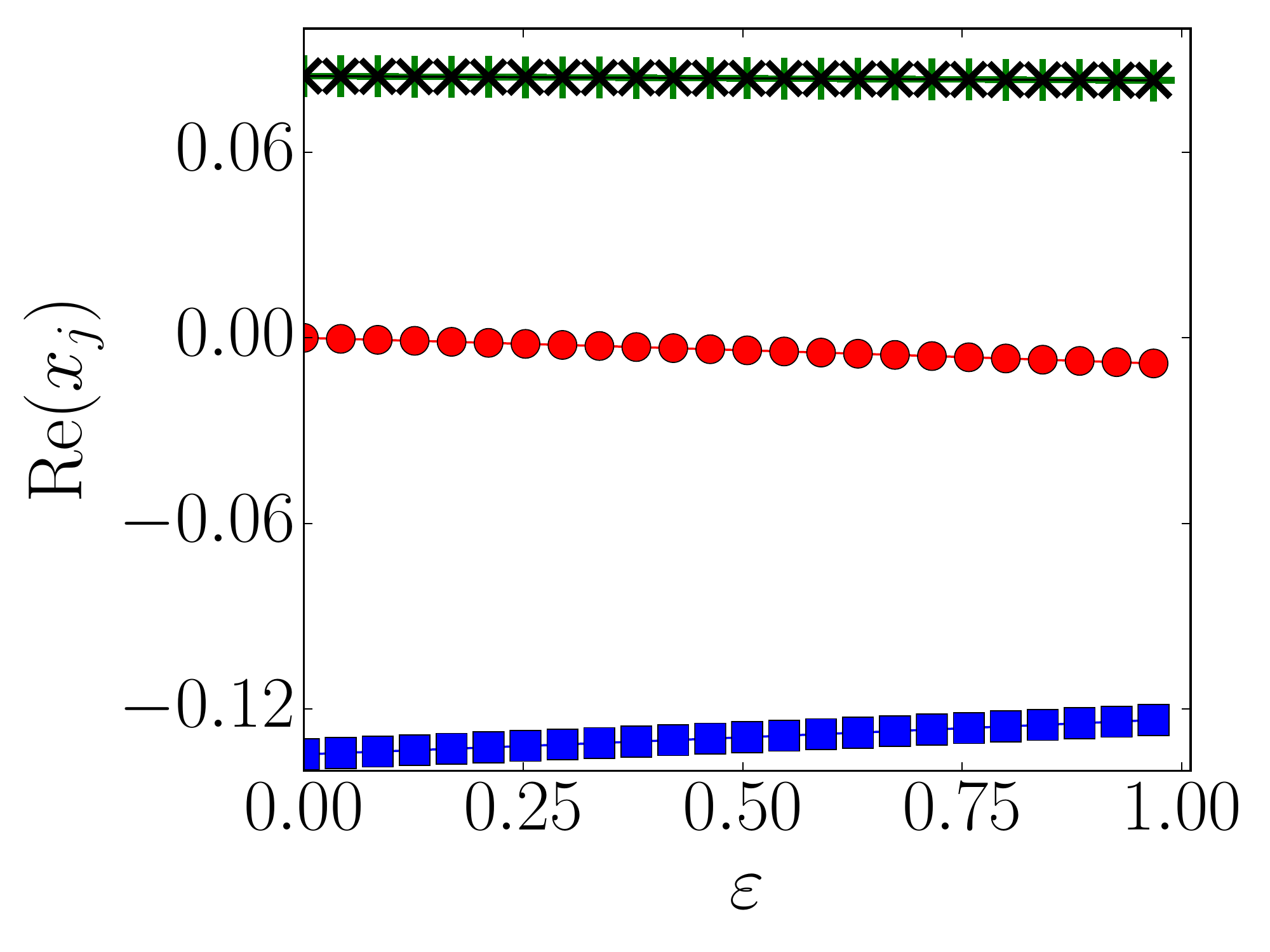}
\caption{\textcolor{black}{Real part $\no{Re}(x_j)$ of four exemplary solutions of Eq.~\eqref{polynomvier}. The physically correct one is selected by changing $\varepsilon$ in Eq.~\eqref{epsilon_method} from zero to one in small steps. At $\varepsilon = 1$ the physical correct solution is the one that originates from $x = 0$ at $\varepsilon = 0$ (marked with red bullets).}}\label{fig:paths_real}%
\end{figure}

In general Eq.~\eqref{polynomvier} has four complex roots \textcolor{black}{$x_j, j\in\{1,...,4\}$}, in practice there are cases with two or all four solutions being real. Only one of the four roots is indeed a physical estimation for $\gamma\sno{EP}$ (and $f\sno{EP}$). The other three (possibly complex) roots arise due to the mathematical structure of approximating the squared energy difference in Eq.~\eqref{2dmatrixentwicklung} up to the second order in $\gamma$ and $f$ and therefore have no physical relevance.
To select the physically correct root out of the four possible ones, it proved to be useful not to set the left hand side of Eq.~\eqref{null} immediately to zero, but to $(1-\varepsilon)\,D$ with an initial value of $\varepsilon = 0$ 
\begin{subequations}
\begin{align}
(1-\varepsilon)\,D\sno{r} &= D\sno{r} + E\sno{r}\,x + F\sno{r}\,y + G\sno{r}\,x^2 + H\sno{r}\,x\,y + I\sno{r}\,y^2,\\
(1-\varepsilon)\,D\sno{i} &= D\sno{i} + E\sno{i}\,x + F\sno{i}\,y + G\sno{i}\,x^2 + H\sno{i}\,x\,y + I\sno{i}\,y^2.
\end{align}
\label{epsilon_method}\noindent
\end{subequations}
Now the $D$ on the left-hand side of Eq.~\eqref{epsilon_method} cancels out that on the right-hand side. Hence, an obvious solution is $x = y = 0$, which corresponds to the centre-point of the octagon. If we assume the coefficients $D$ to $I$ to describe the system at this point most accurately (as they are calculated for an octagon with this centre-point), it is reasonable to treat the root with $x = 0$ at $\varepsilon = 0$ as the distinguished one, which can then be followed by rising $\varepsilon$ in small steps to $\varepsilon = 1$. Now the equation again corresponds to Eq.~\eqref{null}; however, with the difference that the distinguished root resulting from that with $x=0$ at $\varepsilon = 0$ can be taken as the true value of $\gamma\sno{EP}$. The corresponding value for $f\sno{EP}$ results again from Eq.~\eqref{ypsilonvonx}. \textcolor{black}{The approach of selecting the correct solution out of the four possible $x_j$ is illustrated in Fig.~\ref{fig:paths_real} where the paths of the real values of the solutions $x_j$ are plotted as a function of $\varepsilon$. At $\varepsilon = 1$ the physically correct solution is the one that originates from $\no{Re}(x) = 0$ at $\varepsilon = 0$ (marked with red bullets). A plot similar to Fig.~\ref{fig:paths_real} for the imaginary part $\no{Im}(x_j)$ of the four solutions shows qualitatively the same behaviour.}

Note that even in the sometimes occurring case of all four solutions of Eq.~\eqref{null} being complex at $\varepsilon = 1$, it is still possible to make an estimation for the position of the \ac{EP} by taking the estimation for $\gamma\sno{EP}$ and $f\sno{EP}$ at an $\varepsilon$-value smaller than one, where the distinguished root is still real. Sometimes, after a few iteration steps the distinguished solution at $\varepsilon = 1$ becomes real again and in that case the algorithm can converge to an \ac{EP}.

\section*{References}

% \bibliographystyle{unsrt}
% \bibliography{literature}

\end{document}